\documentstyle[12pt]{article}
\newcommand{\Asym}{\mathop{\rm A}}
\newcommand{\Sym}{\mathop{\rm S}}
\newcommand{\bbar}{\overline{\phantom{I}}}

\input epsf.sty

\begin{document}
\begin{titlepage}

\centerline{\large \bf Leading-order analysis of the twist-3 }
\centerline{\large \bf space- and time-like cut vertices in QCD.}

\vspace{10mm}

\centerline{\bf A.V. Belitsky}

\vspace{5mm}

\centerline{\it Bogoliubov Laboratory of Theoretical Physics}
\centerline{\it Joint Institute for Nuclear Research}
\centerline{\it 141980, Dubna, Russia}

\vspace{30mm}

\centerline{Lecture given at the}
\centerline{XXXI PNPI Winter School on Nuclear and Particle Physics}
\centerline{St. Petersburg, Repino, February 24 - March 2, 1997}

\vspace{30mm}

\centerline{\bf Abstract}

\hspace{0.5cm}

We review the recent theoretical progress in the construction
and solution of the evolution equations which govern the scale
dependence for the twist-three structure and fragmentation
functions of the nucleon.

\end{titlepage}

\section{Introduction.}

In the naive parton model \cite{fey71} the cross sections for several
inclusive processes can be expressed in terms of the density of
probability of partons ($\phi$) in a hadron ($h$) $F_{\phi/h}(x)$ or
hadrons in a parton ${\cal F}_{h/\phi}(\zeta)$ with $x$, $1/\zeta$
being the momentum fraction of the final state particle with respect
to the initial one. The probabilistic picture relies on the fact 
that the constituents in the high energy processes behave as a
bunch of noninteracting quanta at small space-time separations.
However, the description of hard reactions using this simple
intuitive picture is very restrictive. The rigorous field
theoretical basis at the leading twist and beyond comes from the
asymptotically free QCD and the use of the factorization theorems
\cite{mue89} which give the possibility to separate the contributions
responsible for physics of large and small distances involved in any
hard reaction. At the lowest twist level, the parton model is
trustworthy and can be mapped onto the language of operator
product expansion approach (OPE) (for deep inelastic scattering).
The contribution of large distances is parametrized by the
distribution (fragmentation) functions mentioned above, or by parton
correlators in a broader sense, which are uncalculable at the moment
from the first principles of the theory. The second ones ---
hard-scattering subprocesses --- can be dealt perturbatively. The
parton distributions are defined in QCD by the target matrix elements
of the light-cone correlators of field operators \cite{col82}. This
representation allows for the estimation of these quantities by the
non-perturbative methods presently available: like the QCD sum rules
\cite{iof88,bra95}, the effective chiral Lagrangians \cite{dia96}, the
MIT bag model \cite{jaf75,jaf90,jaf91,jaf_PR91,sch91,strat93,koi97}.

After addressing for over decades mainly the spin averaged observables
which have probed the nonpolarized substructure of the hadrons and have
yielded an important information on the parton content of nucleon,
the attention has been shifted towards more subtle dynamics underlying
the polarized scattering. A renewed interest in the high energy spin
physics in the last years has been concentrated on the transverse
spin phenomena in hard processes. It revives many ideas developed
over a decade ago. In particular, the notion of the twist-2
transversity distribution $h_1(x)$, first mentioned by Ralston and
Soper \cite{ral79} has been reinvented as well as its twist-3
counterparts have been addressed \cite{art90}. Due to chirality
conservation $h_1(x)$ cannot appear in the inclusive deep inelastic
scattering (DIS) but it can be measured, for instance, in Drell-Yan
reactions through the collision of the transversely polarized
hadrons \cite{ral79,art90} and in semi-inclusive pion production in
DIS on the nucleon \cite{jaf93}. In the last case, it enters into
the cross section as a leading contribution together with the
twist-3 chiral-odd spin-independent fragmentation function
${\cal I}(\zeta)$ \cite{jaf93}. It is well known that there is
considerable difference between the structure and fragmentation
functions. Namely, the moments of the former are expressed in
terms of reduced matrix elements of the tower of the local operators
of definite twist. This property is established by exploiting the
Wilson OPE for inclusive DIS. Although the light-cone expansion for
the fragmentation processes is similar to DIS, the moments of the
corresponding functions are not related to any short-distance limit.
As a substitute for the local operators come the Mueller's time-like
cut vertices \cite{muel78}, which are essentially nonlocal in the
coordinate space so that the analogy to the operator language is
only useful mnemonic.

The transverse spin phenomena \cite{efr95} attained by the inclusive
DIS are associated with ex\-pli\-cit\-ly interaction-dependent,
{\it i.e.} higher twist (twist-3), effects. To disentangle the
underlying complicated dynamics, the unravelling of the twist-3
effects in the hard processes, which manifest the quantum mechanical
interference of partons in the interacting hadrons is needed. At the
twist-3 level, the nucleon has three structure functions $g_2(x)$,
$e(x)$ and $h_L(x)$ and the fragmentation functions corresponding
to each introduced distribution. These new characteristics have been
the subject of intensive theoretical study until recently since
they open a new window to explore the nucleon content. The most
important advantage of the twist-3 structure functions is that
while being important for understanding of the long-range quark-gluon
dynamics they contribute at the leading order in $1/Q$ ($Q$ being the
momentum of the probe) to certain asymmetries \cite{jaf90,jaf91}
and, therefore, are directly accessible by either the polarized
(semi)inclusive DIS \cite{abe96} or Drell-Yan or hadron production
in $e^+e^-$-annihilation processes. To confront the theory
with high-precision data, the knowledge of the size of the logarithmic
violation of the Bjorken scaling by the QCD radiative corrections in
the measurable quantities is highly required.

There exist two equivalent (and complementary) approaches to studying
the $Q^2$-de\-pen\-den\-ce of the structure functions in the leading
logarithmic approximation (LLA) of the perturbation theory. The first
is based on the use of the OPE for the product of currents. It makes
possible the study of logarithmic violation of Bjorken scaling as
well as of the power suppressed contributions (higher twists) responsible
for many subtle phenomena in a polarized scattering. The former is
achieved by the calculation of the anomalous dimensions of the
corresponding local operators. However, there exists an alternative
approach to the analysis of the corresponding quantities which is
based on the evolution equations \cite{lip72,alt77}. In spite of the
fact that the latter has some difficulties as compared to the former,
in the study of higher twists (like the loose of the explicit gauge
and Lorentz invariance of calculations and also the presence of the
overcomplete set of correlation functions) it has an important
advantage as being the closest to the intuitive physical parton-like
picture. There is another advantage of the latter approach to studying
the higher twist effects from the point of view of experimental
capabilities since the OPE provides us with the moments of the structure
functions, and in order to extract the former, one needs to measure the
latter in the whole region of the momentum fraction very accurately.
Obviously, it is a quite difficult task even for the next generation
of colliders. While, with a set of evolution equations at hand, one can
find, in principle, the $Q^2$-dependence of the cross section in
question by putting the experimental cuts on the region of the
attained momentum fractions. This approach can also be used in the
situations when the OPE is no longer valid, {\it i.e.} the time-like
processes.

Thus, the $Q^2$-evolution of the parton distributions \cite{lip72,alt77}
can be predicted un\-am\-bi\-guous\-ly by exploiting the powerful methods
of renormalization group (RG) and QCD perturbation theory. As distinguished
from the leading twist evolution, the twist-3 two-quark frag\-men\-ta\-tion
functions receive contribution from the quark-gluon correlators even
in the limit of asym\-pto\-ti\-cal\-ly large momentum transfer. To solve
the problem, one should correctly account for the mixing of correlators of
the same twist and quantum numbers in the course of renormalization.

In the subsequent discussion we attempt to review the recent theoretical
progress in the study of the twist-3 polarized and nonpolarized
chiral-odd and -even structure and fragmentation functions in the
framework of QCD. The presentation will be organized as follows: The
first chapter is devoted to the consideration of the parton distribution
functions. Here we address the issues of the sum rules, construction of
the basis of independent correlators closed with respect to
renormalization group evolution, the machinery for the calculation
of the evolution kernels, simplification which occurs in the
limit of a large number of colours and the orbital momenta. The last
features make possible the finding of the analytical solution of the
approximated evolution equations. The second chapter concerns the case
of fragmentation functions. We mainly cover the same subjects as in
the first part of the paper.

\section{Space-like cut vertices.}

\subsection{Correlation functions of nonleading twist.}

As we have mentioned in the introduction, the parton distribution
functions in QCD are defined by the Fourier transforms along
the null-plane of the forward matrix element of the parton field
operators product separated by an interval $\lambda$ on the light
cone ($n^2=0$):
\begin{equation}
\label{two-corr}
{\cal F}(\lambda) \equiv {\cal F}(\lambda, 0)
= \phi^* (0) \Phi [0,\lambda n] \phi (\lambda n),
\end{equation}
where $\phi$ denotes a quark $\psi$ or a gluon field $B_\mu$ and
$\Phi$ is a path ordered exponential along the straight line which
insures the gauge invariance of the parton distribution
\begin{equation}
\Phi[{\rm x},{\rm y}]
=P\exp \left(ig({\rm x}-{\rm y})_{\mu}\int_{0}^{1}\!\!d\sigma
B_{\mu}({\rm y}+\sigma ({\rm x}-{\rm y}))\right).
\end{equation}
We suppress the dependence on the renormalization scale $\mu_R$
in Eq. (\ref{two-corr}), necessary to render this quantity well
defined in the field theory. The Fourier transformations from the
coordinate to the momentum space and vice versa are given by
\begin{equation}
\label{Fourier-2}
F(x) = \int \frac{d\lambda}{2\pi} e^{i\lambda x}
\langle h |
{\cal F} (\lambda) | h \rangle ,
\hspace{0.5cm}
\langle h | {\cal F} (\lambda) | h \rangle
= \int dx e^{-i\lambda x} F(x).
\end{equation}
Both of these representations display the complementary aspects
of the factorization. The light-cone position representation is
suitable to make contact with the operator product expansion
approach, while the light-cone fraction representation is
appropriate for es\-tab\-li\-shing the language of the parton model.
Throughout the paper we will use the light-cone position and
the light-cone fraction representations in parallel.

It is well known that in order to endow the theory with parton-like
interpretation and to get much deeper insight into the corresponding
perturbative calculations, it is necessary to use to ghost-free gauges.
Owing to this fact we choose in what follows the light-cone gauge
$B_+ =n^\mu B_\mu = 0$ for the boson field. The advantage of this
gauge is that the gluon field operator $B_\rho$ is related to the
field strength tensor $G_{\rho\sigma}$ via the simple relation
\begin{equation}
\label{B_to_G}
B_{\mu}(\lambda n) =\partial^{-1}_+ G_{+ \mu}(\lambda n)
=\frac{1}{2}\int_{-\infty}^{\infty}dz
\epsilon (\lambda - z)G_{+\mu}(z),\label{G}.
\end{equation}
Here, the residual gauge degrees of freedom are fixed by imposing
antisymmetric boundary conditions on the field, which allows a
unique inversion. Thus, the gauge invariant result can be restored
after all required calculations have been performed.

To trace the origin of the operator definition of the hadron's
parton density we sketch briefly below the factorization procedure
of Ellis-Furmanski-Petronzio (EFP) \cite{EFP83} for the hadron matrix
element of the $T$-product of the electromagnetic currents $T(P,q)$
whose imaginary part defines the familiar structure functions of DIS.
To this end we use the Sudakov decomposition of the four-momentum of
the active parton in transverse and longitudinal components
\begin{equation}
k^\mu = x p^\mu + \alpha n^\mu + k_\perp^\mu.
\end{equation}
Here $n$ is a light-cone vector $n^2 = 0$ normalized with respect
to the four-vector $P = p + \frac{1}{2}M^2n$ of the parent hadron
$h$ of mass $M$, {\it i.e.} $nP=1$, and $p$ is a null vector along the
opposite tangent to the light cone such that $p^2 = 0$, $np = 0$.

The physical observable $T(P,q)$ can be factorized into the hard $H$
and soft $S$ blocks (we omit the Lorentz indices of the currents
$J_\mu$)
\begin{eqnarray}
&&T(P,q)=i \int d^4z e^{i(qz)}
\langle h| T \left\{ J(z)J(0) \right\} | h \rangle
=\int \prod_i d^4 k_i
[H(k_i, q^2) S(k_i, p, {\Lambda}^2)] \nonumber\\
&&=\int \prod_i d^4 k_i dx_i
\delta(x_i-(k_in))[(H(x_ip, q^2)+\dots ) S(k_i, p, {\Lambda}^2)]
\nonumber\\
&&=\int \prod_i dx_i
[(H(x_ip, q^2)+\dots )S({\lambda}_i, p, {\Lambda}^2)]
\equiv [H \otimes S].
\end{eqnarray}
where we have used the collinear expansion of the momentum of the
struck parton with respect to the large +-direction of the hadron
momentum.
\begin{equation}
S(x_i, p, {\Lambda}^2)
=\int \prod_i d^4 k_i
\delta(x_i-(k_in)) S(k_i, p, {\Lambda}^2)
\end{equation}
with
\begin{equation}
S(k_i, p, {\Lambda}^2)=\int \prod_i d^4z_i
e^{\sum_i iz_ik_i}
\langle h |\phi(z_1) \phi(z_2) \dots \phi(z_n)| h \rangle .
\end{equation}
Simple manipulations allow one to write the final answer for the soft
part
\begin{equation}
\label{soft}
S(x_i, p, {\Lambda}^2)
=\int \prod_i \frac{d{\lambda}_i}{2\pi}e^{\sum_i i{\lambda}_ix_i}
\langle h | \phi({\lambda}_1n) \phi({\lambda}_2n)
\dots \phi({\lambda}_nn)|h \rangle .
\end{equation}
Note that $\Phi = 1$ in the gauge we have chosen. By exploiting the
Poincar\'e invariance of the forward matrix element we can exclude
the overall translation and, in a particular case, come to 
Eq. (\ref{two-corr}).

The multiparton distributions corresponding to the interference
of higher Fock com\-po\-nents in the hadron wave functions that emerge
at the twist-3 level are the generalizations of (\ref{two-corr})
to the 3-parton fields and present already in (\ref{soft})
\begin{equation}
\label{three-corr}
{\cal F} (\lambda, \mu) \equiv {\cal F} (\lambda, 0, \mu)
= \phi^* (\mu n) \phi (0) \phi (\lambda n).
\end{equation}
We do not display the quantum numbers of the field operators
since they are not of relevance at the moment. The direct and
inverse Fourier transforms are
\begin{equation}
\label{Fourier-3}
F(x, x') = \int \frac{d\lambda}{2\pi}\frac{d\mu}{2\pi}
e^{i\lambda x - i\mu x'}
\langle h | {\cal F} (\lambda , \mu) | h \rangle ,\hspace{0.5cm}
\langle h | {\cal F}(\lambda, \mu) | h \rangle
= \int dxdx' e^{-i\lambda x + i\mu x'} F(x, x').
\end{equation}
The variables $x$ and $x'$ are the momentum fractions of incoming
$\phi$ and outgoing $\phi^*$ partons, respectively. The
restrictions on their physically allowed values come from the
support properties of the multiparton distribution functions
discussed at length in Ref. \cite{jaf83}, namely $F(x, x')$
vanishes unless $0 \leq x \leq 1$, $0 \leq x' \leq 1$.

Beyond the leading-twist level the intuitive parton-like picture is
not so immediate, as one usually starts with an overcomplete set of
correlation functions. However, the point is that the equations of
motion for field operators imply several relations between
correlators, and the problem of construction of a simpler operator
basis is reduced to an appropriate exploitation of these equalities.
The guiding line to disentangle the twist structure is clearly
seen in the light-cone formalism of Kogut and Soper \cite{kog70}.
Consider, for instance, the correlators containing two quarks
$\bar\psi\psi$. Then, decomposing the Dirac field into ``good"
and ``bad" components with Hermitian projection operators
${\cal P}_{\pm} = \frac{1}{2}\gamma_{\mp}\gamma_{\pm}$:
$\psi_{\pm} = {\cal P}_{\pm} \psi$, we have three possible
combinations $\psi^\dagger_+\psi_+$, $\psi^\dagger_+\psi_-
\pm \psi^\dagger_-\psi_+$, and $\psi^\dagger_-\psi_-$, which are
of twist 2, 3 and 4, respectively. The origin of this
counting lies in the dynamical dependence of the ``bad" components
of the Dirac fermions
\begin{equation}
\label{bad}
\psi_- = -\frac{i}{2}\partial^{-1}_+
\left(
i\not\!\!D_\perp + m
\right) \gamma_+ \psi_+.
\end{equation}
These components depend on the underlying QCD dynamics, i.e.
they implicitly involve extra partons and thus correspond to
the generalized off-shell partons, which carry the transverse
momentum. For this reason we come back to the on-shell massless
collinear partons of the naive parton model, but supplemented with
multiparton correlations through the constraint (\ref{bad}).
The operators constructed of the "good" components
only were named quasi-partonic \cite{lip85}. The advantage of
handling them is that they endow the theory with a parton-like
interpretation for higher twists.

The EFP approach is close to the OPE; this equivalence is established
by identifying the moments of the parton correlation functions
($S$-block) with the reduced matrix elements of the local operators.
The singularities of the product of the currents on the light-cone are
absorbed in the coefficient functions $H$ in front of operators. In the
momentum space they result in the inverse powers of the large momentum
scale $Q$ at which the operators contribute to the cross section and
it is controlled by their twist $\tau$ ($\tau=d_C-s$, where $d_C$ is a
canonical dimension of an operator and $s$ is its Lorentz spin). The
leading contribution comes from the operators of twist $\tau = 2$:
\begin{equation}
\label{twist-2}
\left[ \phi^* (0) \phi (z) \right]^{\rm tw-2}= \sum \frac{1}{n!}
z_{\mu_1} z_{\mu_2} \dots z_{\mu_n}
\left\{
\phi^* (0)
\partial_{\mu_1} \partial_{\mu_2} \dots \partial_{\mu_n}
\phi (0)
-{\rm traces}
\right\}.
\end{equation}
However, as has been established in Refs. \cite{bb88,bal91} one can 
give the definition of the twist without appealing to the concept
of the local operators which is particularly useful in
cases when the short distance expansion is no longer relevant, as
it happens, for instance, for inclusive production of hadron in
the  $e^+e^-$-annihilation {\it et ctr}. The point is that the
nonlocal string operator given by Eq. (\ref{twist-2}) obeys the
Laplace equation
\begin{equation}
\label{Laplaceequation}
\Box \left[ \phi^* (0) \phi (z) \right]^{\rm tw-2} = 0, \mbox{ here }
\Box \equiv \partial^2
\end{equation}
with the boundary condition on the light-cone
\begin{equation}
\phi^* (0) \phi (z) = \sum \frac{1}{n!}
z_{\mu_1} z_{\mu_2} \dots z_{\mu_n}
\left\{
\phi^* (0)
\partial_{\mu_1} \partial_{\mu_2} \dots \partial_{\mu_n}
\phi (0)
\right\}, \mbox{ for } z^2=0.
\end{equation}
The solution can be written in the form
\begin{equation}
\label{Laplacesolotion}
\left[ \phi^* (0) \phi (z) \right]^{\rm tw-2}
= \phi^* (0) \phi (z)
+ \sum_{n=0}^{\infty}
\frac{1}{n!(n-1)!}\left( -\frac{z^2}{4} \right)^n
\int_{0}^{1}dv v^{d/2-1} (v \bar v)^{n-1}
\Box^n \phi^* (0) \phi (vz),
\end{equation}
here $d$ is a dimension of space-time and $\bar v = 1-v$.
For the time-like processes the connection to the local operators
is lost, so that we are left with (\ref{Laplacesolotion}) up to
an arbitrary solution of (\ref{Laplaceequation}) which vanishes
on the light cone. To this accuracy we can define
\begin{equation}
\left[ \phi^* (0) \phi (z) \right]^{\rm tw-2}
= \phi^* (0) \phi (z)
+ \sum_{n=0}^{\infty}
\frac{1}{n!(n-1)!}\left( -\frac{z^2}{4} \right)^n
\int_{1}^{\infty}dv v^{d/2-1} (v \bar v)^{n-1}
\Box^n \phi^* (0) \phi (vz),
\end{equation}
where the integration region mimics the physical domain of the
parton momentum fraction of the annihilation channel and thus the
matrix elements of this string operator possesses the correct
support properties in the light-cone variables.

\subsection{$g_2(x)$.}

Recently, the first experimental data for the measurement of the
transverse spin structure function $g_2(x)$ in the deep inelastic
scattering of the longitudinally polarized muon beam on the
transversely polarized proton target have been reported \cite{abe96}.
Although at present the statistics is too low to be able to extract
its perturbative evolution it proves to be important to know the
theoretical prediction for the latter from QCD. This issue has been
extensively studied in the literature, therefore, we just outline the
main results referring the interested reader to the original works
\cite{sv82,lip84,lip85,rjk66,bb88,ali91,mul96} and reviews
\cite{lip_lect}.

The function we are interested in appears as coefficient in the
Lorentz decomposition of the antisymmetric part of the hadronic
tensor, relevant to the polarized scattering, over the appropriate
tensor structures:
\begin{eqnarray}
W_{[\mu\nu]}
&=& \frac{1}{2\pi}{\rm Im}\ i\!\int dz e^{i(qz)}
\langle h | T\left\{ J_{[\mu} (z) J_{\nu]} (0) \right\} | h \rangle
\nonumber\\
&=& \frac{i}{(pq)}\epsilon_{\mu\nu\rho\sigma}q_\rho
\left\{
s_\sigma g_1(x,Q^2)
+ g_2(x,Q^2)\left( s_\sigma - \frac{(sq)}{(pq)}p_\sigma \right)
\right\}.
\end{eqnarray}

Using the nonlocal light-cone OPE we can express, as has been discussed
in the preceding section, the polarized structure functions in terms
of the hadronic matrix elements of Fourier transformed string
operators
\begin{eqnarray}
S_+g_1(x)&=&\frac{1}{2}\int \frac{d\lambda}{2\pi}e^{i\lambda x}
\langle h |\bar\psi (0)\gamma_+\gamma_5\psi (\lambda n)
| h \rangle,\nonumber\\
\label{g_2}
S^\perp_\sigma g_T(x)&=&
\frac{1}{2}\int \frac{d\lambda}{2\pi}e^{i\lambda x}
\langle h |\bar\psi (0)\gamma^\perp_\sigma \gamma_5\psi (\lambda n)
| h \rangle,
\label{g_1&g_2}
\end{eqnarray}
where $S_\sigma^\perp$ denotes the transverse polarization vector
of the hadron $h$ ($S^2=-M^2$) and $g_T = g_1 + g_2$.

As we have noted above, the higher-twist two-quark operators mix
with multiparton correlators. Moreover, the operator corresponding
to $g_T$ does not possess a definite twist, and as a consequence
could not be renormalized multiplicatively. Taking into account
the equation of motion for the quark field and equality arising from
the use of the Lorentz invariance, one can find \cite{lip84,lip_lect}
\begin{eqnarray}
\label{EOM-g_2}
&&xg_T(x)- \ {\bbar\!\!\!\!\! M}(x) - \ {\bbar\!\!\!\!\! K}(x)
-\int dx' \ {\bbar\!\!\!\! D}(x,x')=0,\\
\label{Lorentz-g_2}
&&xg_1 (x)
=xg_T (x) - x \frac{\partial}{\partial x}\ {\bbar\!\!\!\!\! K}(x)
-x\int dx' \frac{\ {\bbar\!\!\!\! D}(x,x')
+\ {\bbar\!\!\!\! D}(x,x')}{(x' - x)}.
\end{eqnarray}
Here, we have introduced the new correlation functions
\begin{eqnarray}
\label{m-g_2}
&&S^\perp_\sigma\ {\bbar\!\!\!\!\! M}(x)
= \frac{1}{2}\int \frac{d\lambda}{2\pi}
e^{i\lambda x}
\langle h | \bar \psi (0)
m \gamma_+ \gamma^\perp_\sigma \gamma_5
\psi (\lambda n)|h \rangle ,\\
&&S^\perp_\sigma\ {\bbar\!\!\!\!\! K}(x)
= \frac{1}{2}\int \frac{d\lambda}{2\pi}
e^{i\lambda x}
\langle h | \bar \psi (0)
\gamma_+ \partial^\perp_\sigma \gamma_5
\psi (\lambda n)|h \rangle , \\
&&S^\perp_\sigma\ {\bbar\!\!\!\! D}_1(x,x')
= \frac{1}{2}\int \frac{d\lambda}{2\pi} \frac{d\mu}{2\pi}
e^{i\lambda x - i\mu x'}
\langle h | \bar \psi (\mu n)
{\rm g} \gamma_+ \!\not\!\! B^\perp (0) \gamma^\perp_\sigma \gamma_5
\psi (\lambda n)|h \rangle ,\\
\label{D-g_2}
&&S^\perp_\sigma\ {\bbar\!\!\!\! D}_2(x',x)
= \frac{1}{2}\int \frac{d\lambda}{2\pi} \frac{d\mu}{2\pi}
e^{i\mu x' - i\lambda x}
\langle h | \bar \psi (\lambda n)
{\rm g} \gamma_+ \gamma^\perp_\sigma \!\not\!\! B^\perp (0) \gamma_5
\psi (\mu n)|h \rangle ,
\end{eqnarray}
and
\begin{equation}
\ {\bbar\!\!\!\! D} (x,x')
=\frac{1}{2}
\left[ \ {\bbar\!\!\!\! D}_1 (x,x')
+ \ {\bbar\!\!\!\! D}_2 (x',x) \right]
\end{equation}
is a $C$-even combination of correlators which can enter into
the cross section due to even photon state in the $t$-channel
under the charge conjugation . The derivative in the correlation
function $\ {\bbar\!\!\!\!\! K}(x)$ acts on the quark field before
setting its argument on the light cone.

Solving the system of the differential equations (\ref{EOM-g_2})
and (\ref{Lorentz-g_2}) with respect to $g_T(x)$ the integration
constant can be found from the support properties of the
distribution: $g_T(x) = 0$ for $|x|\geq 1$. The solution provides
us with the following relation between these functions:
\begin{eqnarray}
\label{SR-g_2}
g_T(x) &=& \int_{x}^{1} \frac{d\beta}{\beta}g_1(\beta)
+ \frac{1}{x}\ {\bbar\!\!\!\!\! M}(x)
- \int_{x}^{1} \frac{d\beta}{\beta^2}\ {\bbar\!\!\!\!\! M}(\beta)
\nonumber\\
&+& \int_{x}^{1} \frac{d\beta}{\beta}\int
\frac{d\beta'}{\beta' - \beta}
\left[
\frac{\partial}{\partial \beta} Y (\beta, \beta')
+\frac{\partial}{\partial \beta'} Y (\beta', \beta)
\right],
\end{eqnarray}
where
\begin{equation}
Y(x,x') = (x - x')\ {\bbar\!\!\!\! D}(x, x').
\end{equation}
Here $Y(x,x')$ is explicitly gauge invariant distribution since
$\ {\bbar\!\!\!\! D}$ is gauge variant provided we use a gauge
other than the light-cone. To see this, we exploit the advantages of
the light-cone gauge, where the gluon field is expressed in terms
of the field strength tensor by Eq. (\ref{B_to_G}) and take into
account the relation
\begin{equation}
\frac{1}{2} \int \frac{d\lambda}{2\pi}
e^{\pm i \lambda x} \epsilon (\lambda - z)
= \pm \frac{i}{2\pi} {\rm PV} \frac{1}{x} e^{\pm i z x},
\end{equation}
we can easily obtain the expressions of the gauge-invariant
quantities in terms of three-particle string operators which are
nonlocal generalization of the Shuryak-Vainstein operators
${^\pm\!{\cal S}_\sigma}$ \cite{sv82}. Generically
\begin{eqnarray}
S^\perp_\sigma Y_1(x, x')
&=& \frac{1}{2}\int \frac{d\lambda}{2\pi}
e^{i\lambda x - i\mu x'}
\langle h | {^+\!{\cal S}_\sigma} (\lambda, 0, \mu)
|h \rangle ,\\
S^\perp_\sigma Y_2(x', x)
&=& \frac{1}{2}\int \frac{d\lambda}{2\pi}
e^{i\mu x' - i\lambda x}
\langle h | {^-\! {\cal S}_\sigma} (\mu, 0, \lambda)
|h \rangle ,
\end{eqnarray}
with
\begin{eqnarray}
{^\pm\!{\cal S}_\sigma}(\lambda, 0, \mu)
&=& \bar{\psi}(\mu n) i {\rm g} \gamma_+
\left[
i \widetilde G^\perp_{\sigma +}(0)
\pm
\gamma_5 G^\perp_{\sigma +}(0)
\right]
\psi(\lambda n),
\end{eqnarray}
where $\widetilde G_{\mu\nu} = \frac{1}{2}\epsilon_{\mu\nu\rho\sigma}
G_{\rho\sigma}$ is the dual field strength tensor and we have used the
relation $\epsilon^\perp_{\rho\sigma}G^\perp_{\rho +} = \widetilde
G^\perp_{\sigma +}$ with the two-dimensional antisymmetric tensor
$\epsilon^\perp_{\rho\sigma} \equiv \epsilon^\perp_{\rho\sigma + -}$.

Thus, the leading order analysis (\ref{SR-g_2}) suggests that the
structure function can be written as the following sum:
\begin{equation}
g_2(x) = g_2^{WW} (x) + \widetilde g_2(x),
\end{equation}
where
\begin{equation}
g_2^{WW}(x) = - g_1(x) + \int_{x}^{1}\frac{d\beta}{\beta}g_1(\beta)
\end{equation}
is the twist-2 Wandzura-Wilczek contribution to the structure
functions \cite{ww77}, while $\widetilde g_2(x)$ is a genuine
twist-3 (explicitly interaction-dependent up to unessential
quark-mass kinematical contribution) part which is expressed
via the integral of the matrix element of the nonlocal operators
which measure the quark-gluon correlation function in the target
nucleon.

The distribution functions defined by Eqs. (\ref{g_2}),
(\ref{m-g_2})-(\ref{D-g_2}) form the redundant basis of operators
closed under renormalization group evolution. Going over to the
operators composed of the "good" components only, we are forced to
consider the renormalization of the correlators\footnote{In this
discussion, we restrict ourselves to consideration of the
non-singlet channel only.} $ {\bbar\!\!\!\!\! M}(x)$ and $Y(x,x')$.

\subsection{$e(x)$.}
\label{e-distribution}

In the unpolarized case we define the following redundant basis of
the chiral-odd twist-3 correlations:
\begin{eqnarray}
\label{DefUnpoCF-e}
&&\hspace{-0.7cm}e(x)
=\frac{x}{2}\int \frac{d\lambda}{2\pi}
e^{i\lambda x}
\langle h | \bar \psi (0)  \psi (\lambda n)|h \rangle ,
\\
\label{DefUnpoCF-M}
&&\hspace{-0.7cm}M(x)
=\frac{1}{2}\int \frac{d\lambda}{2\pi}
e^{i\lambda x}
\langle h | \bar \psi (0) m\gamma_+  \psi (\lambda n)|h \rangle ,
\\
\label{DefUnpoCF-D1}
&&\hspace{-0.7cm}D_1 (x,x')
=\frac{1}{2}\int \frac{d\lambda}{2\pi} \frac{d\mu}{2\pi}
e^{i\lambda x-i\mu x'}
\langle h | \bar \psi (\mu n)
{\rm g} \gamma_+\! \not\!\! B^\perp(0)
\psi (\lambda n)|h \rangle ,
\\
\label{DefUnpoCF-D2}
&&\hspace{-0.7cm}D_2 (x',x)
=\frac{1}{2}\int \frac{d\lambda}{2\pi} \frac{d\mu}{2\pi}
e^{i\mu x' - i\lambda x}
\langle h | \bar \psi (\lambda n)
{\rm g}\! \not\!\! B^\perp(0) \gamma_+
\psi (\mu n)|h \rangle.
\end{eqnarray}
The functions $D_1$ and $D_2$ are related by complex conjugation
$\left[ D_1 (x,x') \right]^* = D_2 (x',x)$. The quantities
determined by these equations form a closed set under
renormalization; however, they are not independent, since there
is a relation between them due to the equation of motion for the
Heisenberg fermion field operator:
\begin{equation}
\label{EOM-e}
e(x)-M(x)-\int dx' D(x,x')=0,
\end{equation}
where again we have introduced the convention
\begin{eqnarray}
&&D (x,x')
=\frac{1}{2}
\left[ D_1 (x,x') + D_2 (x',x) \right].
\end{eqnarray}
This function is real-valued and antisymmetric with respect to the
exchange of its arguments:
\begin{eqnarray}
\left[D(x,x') \right]^* = D(x,x') , \qquad  D (x,x') = - D (x',x).
\end{eqnarray}
Below, in section \ref{abel-evol}, as an illustration of the
self-consistency of the whole approach to the higher twists we
present a set of coupled RG equations for the correlation functions
determined by Eqs. (\ref{DefUnpoCF-e})-(\ref{DefUnpoCF-D2}) derived
in the abelian gauge theory. Relation (\ref{EOM-e}) provides
a strong check of our calculations\footnote{This fact follows from
general renormalization properties of gauge-invariant operators
as one expects that the counter term for the equation of motion
operator can be given only by the operator itself. Its matrix
element, being taken with respect to the physical
state, decouples completely from the renormalization group
evolution.}. It allows the reduction, as we have mentioned above,
of the RG analysis to the study of scale dependence of the
three-parton $D$ and mass-dependent $M$ correlators only.

Introducing the quantity
\begin{eqnarray}
\label{GIDefUnpoCF}
Z(x,x') = (x-x')D(x,x').
\end{eqnarray}
we can easily obtain from Eqs. (\ref{DefUnpoCF-D1}) and
(\ref{DefUnpoCF-D2}) the definition of the gauge-invariant
quantities $Z$ in terms of three-particle string operators,
namely
\begin{eqnarray}
\label{DefUnpolLRO-Z}
Z(x,x') =
\frac{1}{2}
\int \frac{d\lambda}{2\pi}\frac{d\mu}{2\pi}
e^{i\lambda x - i\mu x'}
\langle h |
{\cal Z}(\lambda,\mu) + {\cal Z}(-\mu,-\lambda)
| h \rangle,
\end{eqnarray}
where
\begin{eqnarray}
\label{DefLRO-Z}
{\cal Z}(\lambda , \mu) \equiv {\cal Z}(\lambda ,0, \mu)
= \frac{1}{2}
\bar \psi (\mu n) {\rm g} G_{+ \rho} (0)
\sigma^\perp_{\rho +} \psi (\lambda n).
\end{eqnarray}
In the same way, for a mass-dependent non-local string operator
\begin{eqnarray}
\label{DefLRO-M}
{\cal M}^j(\lambda) \equiv {\cal M}^j(\lambda, 0)
=\frac{m}{2}\bar \psi (0) \gamma_+ (iD_+ (\lambda) )^j
\psi (\lambda n),
\end{eqnarray}
the Fourier transform is
\begin{eqnarray}
\label{DefUnpolLRO-M}
M^j (x) &=& x^j M (x)
= \int \frac{d\lambda}{2\pi}
e^{i\lambda x}
\langle h |{\cal M}^j( \lambda ) |h \rangle .
\end{eqnarray}

For the spin-dependent scattering discussed below, the only
difference is that one should insert also a $\gamma_5$-matrix
between the fields in the definitions of the string operators
(\ref{DefLRO-Z}), (\ref{DefLRO-M}).

\subsection{$h_L(x)$.}

Analogously, the set of correlation functions for the polarized
case is as follows:
\begin{eqnarray}
&&\hspace{-0.7cm}h_1(x) \label{DefPoCF-h_1}
=\frac{1}{2}S_\sigma^\perp \int \frac{d\lambda}{2\pi}
e^{i\lambda x}
\langle h | \bar \psi (0)i \sigma^{\ \,\perp}_{+ \sigma} \gamma_5
\psi (\lambda n)|h \rangle ,\\
&&\hspace{-0.7cm}h_L(x) \label{h_L}
=\frac{x}{2}\int \frac{d\lambda}{2\pi}
e^{i\lambda x}
\langle h | \bar \psi (0)i \sigma_{+ -} \gamma_5
\psi (\lambda n)|h \rangle ,\\
&&\hspace{-0.7cm}\widetilde M(x)
=\frac{1}{2}\int \frac{d\lambda}{2\pi}
e^{i\lambda x}
\langle h | \bar \psi (0) m\gamma_+ \gamma_5
\psi (\lambda n)|h \rangle ,\\
&&\hspace{-0.7cm}K(x) \label{K}
=\frac{1}{2}\int \frac{d\lambda}{2\pi}
e^{i\lambda x}
\langle h | \bar \psi (0) i\gamma_+\!\not\!\partial_\perp
\gamma_5
\psi (\lambda n)|h \rangle ,\\
&&\hspace{-0.7cm}\widetilde D_1 (x,x')
=\frac{1}{2}\int \frac{d\lambda}{2\pi} \frac{d\mu}{2\pi}
e^{i\lambda x-i\mu x'}
\langle h | \bar \psi (\mu n)
{\rm g} \gamma_+\! \not\!\! B^\perp(0) \gamma_5
\psi (\lambda n)|h \rangle ,\\
&&\hspace{-0.7cm\widetilde }D_2 (x',x) \label{DefPoCF-D2}
=\frac{1}{2}\int \frac{d\lambda}{2\pi} \frac{d\mu}{2\pi}
e^{i\mu x' - i\lambda x}
\langle h | \bar \psi (\lambda n)
{\rm g} \gamma_+\! \not\!\! B^\perp(0) \gamma_5
\psi (\mu n)|h \rangle,
\end{eqnarray}

Besides the identity arising from the equation of motion
\begin{equation}
\label{EOM-h_L}
h_L(x)-\widetilde M(x) - K(x)
-\int dx'\widetilde D(x,x')=0,
\end{equation}
there is an equation provided by the Lorentz invariance
\begin{equation}
\label{tw-2-3}
2xh_1 (x)
=2h_L (x) - x \frac{\partial}{\partial x}K(x)
-2x\int dx' \frac{\widetilde D(x,x')}{(x' - x)}.
\end{equation}
It means that both parts of this equality are expressed in terms
of matrix elements of different components of one and the same
twist-2 tensor operator, and thus should possess the same anomalous
dimensions. Again, we have introduced the $C$-even quantity
$\widetilde D$, which has the properties
\begin{equation}
\left[ \widetilde D (x,x') \right]^* = \widetilde D (x, x'),
\quad
\widetilde D (x,x') = \widetilde D (x',x).
\end{equation}
Following the same line as above, we come to the equation
\begin{eqnarray}
\label{sumrule}
h_L(x)
&=& 2x^2 \int_{x}^{1}\frac{d\beta}{\beta^2} h_1(\beta)
+ \widetilde M (x)
- 2x^2\int_{x}^{1}\frac{d\beta}{\beta^3} \widetilde M (\beta)
\nonumber\\
&+& x^2 \int_{x}^{1} \frac{d\beta}{\beta^2}
\int \frac{d\beta'}{\beta' - \beta}
\left[ \frac{\partial}{\partial \beta}
- \frac{\partial}{\partial \beta'} \right]
\widetilde Z (\beta, \beta').
\end{eqnarray}
A similar relation was found by Jaffe and
Ji \footnote{The corresponding expressions in Ref. \cite{jaf91}
contain misprints.} in Ref. \cite{jaf91}. Here
the dynamical twist-3 contribution is explicitly related to
a particular integral of the three-parton correlation function
$\widetilde Z$. In terms of local operators it looks like
\begin{equation}
(n + 3)[h_L]_n = 2 [h_1]_{n+1} + (n + 1) \widetilde M_n
+\sum_{l=1}^{n} (n-l+1) \widetilde Z_n^l ,
\end{equation}
and the definition of moments of distribution functions is given
by Eq. (\ref{mom-2-3}).

As before, excluding the functions (\ref{h_L}) and (\ref{K}), and using
the relations (\ref{EOM-h_L}) and (\ref{tw-2-3}), we can chose the basis
of independent functions in the form: $h_1(x)$, $\widetilde M(x)$,
$\widetilde D(x, x')$.

\subsection{Construction of the evolution equations.}

As long as $z^2 \neq 0$, the renormalization of the $T$-product of
the operators $T\left\{ \phi^*(0) \phi(z)\right\}$ is trivial and
reduced to the familiar renormalization constants $Z$ of the
fundamental field operators entering into the Lagrangian density.
However, if $z=0$ or $z^2=0$ an additional divergence enters the game.
This is a well-known fact from the renormalization theory since the
product of (at least) two field operators entering into the same
space-time point (or on the light cone) produces an ill-defined
quantity from the point of view of the theory of distributions
and the corresponding infinities have to be regularized and subtracted.
In the momentum space this results in ultraviolet (UV) divergences
of the momentum integrals in the perturbation theory, and as a
by-product this causes the logarithmic dependence of the parton
densities on the normalization point. This dependence is governed
by the renormalization group. The evolution equations for the leading
twist correlation functions determining their $Q^2$ dependence
can be interpreted in terms of the kinetic equilibrium of partons
inside a hadron (for distribution functions) or hadrons inside a
parton (for fragmentation function) under the variation of the
ultraviolet transverse momentum cut-off \cite{lip72}. However,
beyond the leading twist the probabilistic picture is lost due
to a quantum mechanical interference and more general quantities
emerge, {\it i.e.} multiparticle parton cor\-re\-la\-tion functions,
whose scale dependence is determined by the Faddeev-type evolution
equation with a pair-wise particle interaction \cite{lip85}.

\begin{figure}[htb]
\mbox{
\hspace{5.3cm}
\epsffile{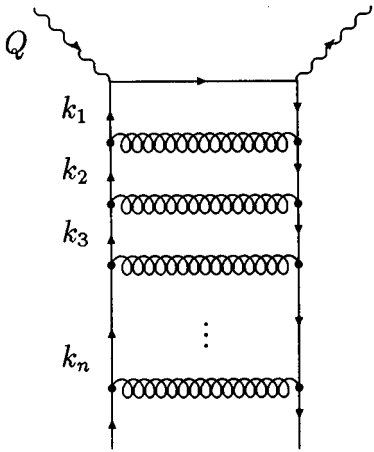}}
\vspace{0.1cm}
{\caption{\label{ladder}
Ladder diagram for deep inelastic scattering.
}}
\end{figure}

There are two sources of the logarithmic dependence of the
correlation functions. The first is the divergences of the
transverse momentum integrals of the particles interacting
with the vertex and forming the perturbative loop. Another
source is the divergences due to the virtual radiative
corrections. In the renormalizable field theory the latter
are factorized into the renormalization constants $Z$ of the
corresponding Green functions. However, owing to the specific
features of the renormalization in the light-like gauge,
extensively reviewed in the next section, there is mixing
of correlation functions due to the renormalization of field
operators. The latter fact is closely related to the matrix
nature of renormalization constants of the elementary Green
functions in the axial gauge. For example, after renormalization
of the fermionic propagator the matrix structure of the bare
vertex could be changed, in general, since the renormalization
matrix acts on the spinor indices of the vertex (see Eq.
(\ref{renorm})).

In the Leading Logarithmic Approximation (LLA) there is a strong
ordering \cite{lip72,grlip72_2} of transverse particle momenta as
well as their minus components, so that only the particles entering
into the divergent virtual block $\Sigma,\Gamma,\Pi$ or the
particles adjacent to the vertex can achieve the maximum values of
$|k_\perp |$ and $\alpha$ (see Fig.~\ref{ladder}):
\begin{eqnarray}
&|k_{n \perp}| \ll ... \ll
|k_{2 \perp}| \ll |k_{1 \perp}| \ll \Lambda ,& \nonumber\\
&\alpha_n \ll ... \ll \alpha_2 \ll \alpha_1 ,&
\end{eqnarray}
while the plus components of the parton momenta are of the same
order of magnitude
\begin{equation}
x_n \sim ... \sim x_2 \sim x_1
\end{equation}
for a $n$-rank ladder-type diagram.

\subsection{Renormalization in the light-cone gauge.}
\label{renormalization}

A peculiar feature of the light-like gauge is the presence of
the spurious IR pole $1/k_+$ in the density matrix of the gluon
propagator
\begin{eqnarray}
D_{\mu\nu}(k)=\frac{d_{\mu\nu}(k)}{k^2+i0}, \quad
d_{\mu\nu}= g_{\mu\nu}-\frac{k_\mu n_\nu + k_\nu n_\mu}{k_+}.
\end{eqnarray}
The central question is how to handle this unphysical pole when
$k_+ = 0$. There are two different ways to treat it which employ the
Cauchy principal value (PV) and Mandelstam-Leibbrandt prescriptions
(ML) \cite{lei87}:
\begin{eqnarray}
{\rm PV}\frac{1}{k_+} &=& \frac{1}{2}
\left\{ \frac{1}{(kn) + i0} + \frac{1}{(kn) - i0} \right\}, \\
{\rm ML}\frac{1}{k_+} &=& \frac{(kn^*)}{(kn)(kn^*) + i0},
\end{eqnarray}
with an arbitrary four-vector $n^*$ satisfying $n^{*2} = 0$,
$nn^* = 1$ (without loss of generality, we can put
it equal to $p$).

Here, we outline the one-loop renormalization program for the abelian
gauge theory with PV prescription. In the next section, we show, using 
a simple example, the difference one encounters in dealing with the
ML prescription.

Due to an additional power of the transverse momentum $k_\perp$ in
the numerator of the density matrix of the gluon propagator, there
exist extra UV divergences of the Feynman graphs which are
absent in the usual isotropic gauges. For our practical aims, we
limit ourselves to the calculation of the one-loop expressions
for the propagators and vertex functions. This is sufficient for
reconstruction of the equations in LLA using the re\-nor\-ma\-li\-za\-tion
group invariance.

The unrenormalized fermion Green function is given by the expression
\begin{equation}
G^{-1}(k)=\not\! k-m_0-\Sigma (k),
\end{equation}
where $\Sigma (k)$ is a self-energy operator. Calculating the latter to
the one-loop accuracy we get the following result
\begin{equation}
G(k)=(1-\Sigma_1)U_2^{-1}(k)\frac{1}{\not\! k-m}U_1(k),
\end{equation}
where
\begin{eqnarray}
&&U_1 (k) =1-\frac{m}{k_+}\Sigma_2(k)\gamma_+
-\frac{1}{k_+}(\Sigma_2(k)-\Sigma_1)\gamma_+ \!\!\!\not\! k,\\
&&U_2 (k) =1+\frac{m}{k_+}\Sigma_2(k)\gamma_+
+\frac{1}{k_+}(\Sigma_2(k)-\Sigma_1)\not\! k \gamma_+,
\end{eqnarray}
and
\begin{equation}
\Sigma_1 =\frac{\alpha}{4\pi}\ln \Lambda^2,\hspace{0.3cm}
\Sigma_2(k) =\frac{\alpha}{4\pi}\ln \Lambda^2
\int dz' \frac{z}{(z-z')}\Theta_{11}^0 (z',z'-z).
\end{equation}
Here $m$ is a renormalized fermion mass related to the bare
quantity by the well-known relation
\begin{equation}
m_0 =m (1-3\Sigma_1).
\end{equation}
The functions $\Theta^m_{i_1 i_2 ... i_n }$ used throughout the paper
are given by the formula
\begin{equation}
\label{Theta}
\Theta^{m}_{i_1 i_2 ... i_n}
(x_1,x_2,...,x_n)=\int_{-\infty}^{\infty}\frac{d\alpha}{2\pi i}
\alpha^m \prod_{k=1}^{n}\left(\alpha x_k -1 +i0 \right)^{-i_k}.
\end{equation}
For our purposes it is sufficient to have an explicit form of the
simplest functions
\begin{eqnarray}
&&\Theta^0_1 (x) = 0,\\
&&\Theta^0_2 (x) = \delta (x),\\
&&\Theta^0_{11} (x_1,x_2)
=\frac{\theta(x_1)\theta(-x_2)
-\theta(x_2)\theta(-x_1)}{x_1-x_2},
\end{eqnarray}
since the others can be expressed in their terms via the following
relations:
\begin{eqnarray}
&&\Theta^0_{21} (x_1,x_2)
=\frac{x_2}{x_1-x_2}\Theta^0_{11} (x_1,x_2),\\
&&\Theta^1_{21} (x_1,x_2)
=\frac{1}{x_1-x_2}\Theta^0_{11} (x_1,x_2)
- \frac{1}{x_1-x_2}\Theta^0_2 (x_1),\\
&&\Theta^0_{22} (x_1,x_2)
= -\frac{2x_1x_2}{(x_1-x_2)^2} \Theta^0_{11} (x_1,x_2),\\
&&\Theta^0_{111} (x_1,x_2,x_3)
=\frac{x_2}{x_1-x_2}\Theta^0_{11} (x_2,x_3)
-\frac{x_1}{x_1-x_2}\Theta^0_{11} (x_1,x_3),\\
&&\Theta^1_{111} (x_1,x_2,x_3)
=\frac{1}{x_1-x_2}\Theta^0_{11} (x_2,x_3)
-\frac{1}{x_1-x_2}\Theta^0_{11} (x_1,x_3).
\end{eqnarray}

As we can easily observe, the renormalization constants are not
numbers any more but matrices acting on the spinor indices of fermion
field operators. Moreover, the re\-nor\-ma\-li\-za\-tion constants
depend on the fractions $k_{i+}$. The origin of this can be traced back
to the lack of the rescaling invariance
\begin{equation}
d_{\mu\nu}(\rho k) \neq d_{\mu\nu}(k) ,
\end{equation}
obeyed by the unregularized propagator.

An abelian Ward identity leads to the equality of the renormalization
constants of the gauge boson wave-function and a charge $Z_3 =
Z_{\rm g}$ so that the corresponding logarithmic dependence on the UV
cut-off cancels in the sum of these two contributions in the evolution
equation and, therefore, we can neglect the fermion loop insertions
into the boson line (in the QCD case this is no longer true).

One can easily calculate the vertex function to the same accuracy.
The result is
\begin{equation}
\Gamma_\rho (k_1,k_2)=(1+\Sigma_1)U_1^{-1}(k_1){\cal G}_\rho U_2(k_2),
\end{equation}
where
\begin{equation}
{\cal G}_\rho
=\gamma_\rho
-(\not\! k_1 -m){\cal Q}_\rho(k_1,k_2) \gamma_+
-\gamma_+ {\cal Q}_\rho(k_1,k_2) (\not\! k_2 -m)
\end{equation}
and
\begin{equation}
{\cal Q}_\rho(k_1,k_2)
=\Sigma_3 (k_1) \gamma_- \gamma_+ \gamma_\rho^\perp
+\Sigma_3 (k_2) \gamma_\rho^\perp  \gamma_+ \gamma_-,
\end{equation}
here
\begin{equation}
\Sigma_3 (k_i)
=\frac{\alpha}{8\pi}\ln \Lambda^2
\int dz' \frac{(z_i-z')}{z'}\Theta_{111}^1 (z',z'-z_1,z'-z_2).
\end{equation}
Apart from the graphs we accounted for, there exists an additional
UV divergence of the virtual Compton scattering amplitude; however,
we do not need its explicit expression for our practical purposes.
This completes the consideration of virtual corrections which cause
the logarithmic dependence on the UV momentum cut-off of the
quantities in question.

\subsection{Difference between the PV and ML prescriptions.}

In the subsequent discussion we will use both the prescriptions for
the gluon propagator in our practical calculations. The first
one (PV) will be used in the momentum space
\cite{lip84,lip85,bel96,bel97}, while the second one in the
coordinate space formulation \cite{mul96,bel97}. As a by-product
we verify that both of them do lead to the same result. However,
it is worthwhile to realize the distinctive features one faces in
the computation of the same quantities.

The most important difference of the second prescription is the
presence of the additional absorptive part \cite{bas96} of the
vector boson Green function, namely
\begin{equation}
{\rm Disc}\left\{ {\rm ML} \frac{d_{\mu\nu}(k)}{k^2+i0} \right\}
= - 2 \pi i \theta(k_+)
\left[
d_{\mu\nu}(k) \delta(k^2)
- \frac{k_-}{k^2} \left( k_\mu n_\nu + k_\nu n_\mu \right)
\delta (k_+ k_-)
\right].
\end{equation}
The second contribution is of the "ghost" type since it has the
wrong sign as compared to the conventional one. However, it is not
an optional choice but it is an unavoidable consequence of equal time
canonical quantization \cite{lei87}. The consequence of this
addendum can be easily recovered in the calculation of the one-loop
evolution kernels. Let us consider, for instance, the
coordinate space formalism \cite{mul96,bel97}. Then, the operator
vertices $V(\lambda_j)$ have only exponential dependence on the
position on the light-cone
\begin{equation}
V(\lambda_j) = \Gamma_V e^{i\sum_j\lambda_j {k_j}_+ }.
\end{equation}
A simple calculation of the one-loop diagram for the leading twist
density with $\Gamma_V = \gamma_+$ gives
\begin{equation}
\left[ \gamma_+e^{i\lambda k_+} \right]_{\Lambda^2}
= \frac{\alpha}{2\pi} \ln \Lambda^2
\int_{0}^{1}dy \left[ \gamma_+e^{iy\lambda k_+} \right]
\left\{ \left[\frac{2}{\bar y}\right]_+ - 1 - y \right\},
\end{equation}
so that the well-defined $1/\bar y_+$-distribution appears already
in the contribution with "real-emission". The self-energy insertions
into external legs produce a familiar $\frac{\alpha}{2\pi}
\frac{3}{2}\delta(\bar y)$ term. Contrary to this, as we have seen
in section~\ref{renormalization}, the renormalization constants of
the field operators in the light-cone gauge with the PV prescription 
turn out to be momentum dependent, and as a by-product the 
plus-prescription fulfilling occurs only in the sum of the real and 
virtual corrections.

\subsection{Evolution equations in the abelian gauge theory.}
\label{abel-evol}

In this section, we present a pedagogical illustration of the
renormalization group mixing problem for the redundant basis
of unpolarized correlation functions defined by Eqs.
(\ref{DefUnpoCF-e})-(\ref{DefUnpoCF-D2}), in the framework of
the abelian gauge theory. Our aim here is to show the self-consistency
of the whole approach we have used since the equations derived below
satisfy the constraint equality given by Eq. (\ref{EOM-e}), which
are further employed to reduce the overcomplete set of correlators
to the basis of independent functions.

\begin{figure}[htb]
\mbox{
\hspace{2.6cm}
\epsffile{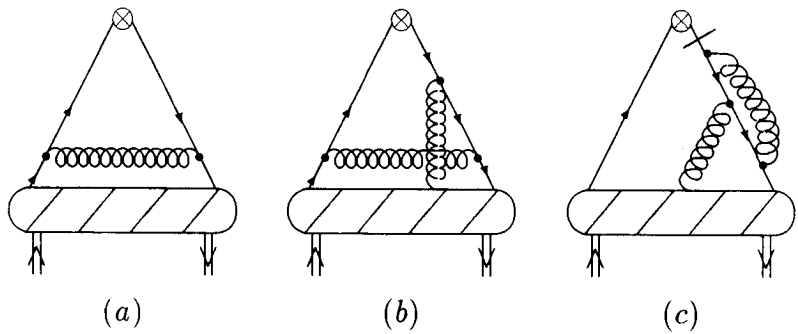}}
\vspace{0.1cm}
{\caption{\label{two-part}
One-loop radiative corrections to the
two-particle correlators in the abelian gauge theory. The fermion
propagator crossed with a bar on diagram (c) means the contraction
of the corresponding line into the point.
}}
\end{figure}

\begin{figure}[htb]
\mbox{
\hspace{0.6cm}
\epsffile{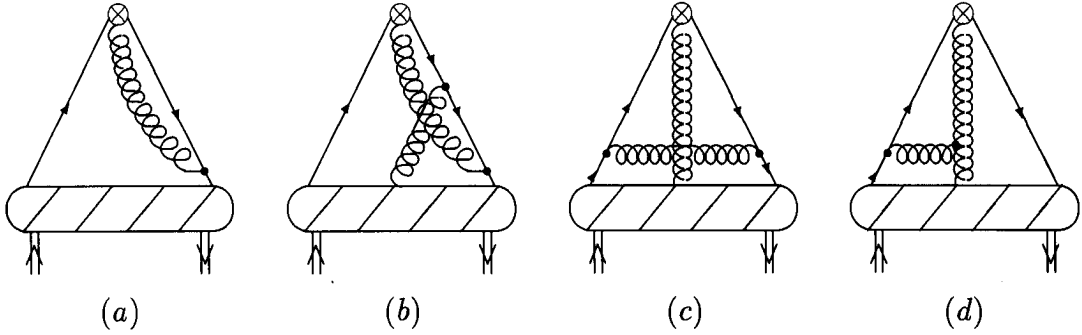}}
\vspace{0.1cm}
{\caption{\label{three-pa}
The one-loop renormalization of the three-parton correlation
functions. Self-energy insertions into external legs are implied.
}}
\end{figure}

The one-loop Feynman diagrams giving rise to the transition
amplitudes of two-particle correlation functions into the two-
and three-parton ones are shown in Fig.~\ref{two-part} (a,b).
The last figure (c) on this picture is specific of the vertices
having non-quasi-partonic form \cite{lip85}, that is for $e(x)$;
it displays the addendum due to the contact term that results from
the cancellation of the propagator adjacent to the quark-gluon and
bare vertices. As an output the vertex acquires the three-particle
piece. The radiative corrections to the three-parton correlators
are presented in Fig.~\ref{three-pa} (a,b,c).

A straightforward calculation yields the evolution equations
for the spin-independent case in the form \cite{bel97}
\begin{eqnarray}
\dot M(x)
&=& - \frac{\alpha}{2\pi}\int d \beta M(\beta)
\left\{
2
\left[
\frac{ \beta }{( x - \beta )}\Theta_{11}^0 (x, x - \beta)
\right]_+
+
\frac{\beta + x}{\beta} \Theta_{11}^0 (x, x-\beta)
\right\},\\
&&\nonumber\\
\dot e(x)
&=&\frac{\alpha}{2\pi}\int d\beta
\Biggl(
e (\beta )
\left\{
\frac{x}{\beta}\Theta_{11}^0 (x, x - \beta)
+\frac{1}{2} \delta (\beta - x)
\right\} \nonumber\\
&-&
M (\beta)
\left\{
2
\left[
\frac{ \beta }{( x - \beta )}
\Theta_{11}^0 (x, x - \beta)\right]_+
+
x \Theta_{21}^1 (x, x - \beta)
+
2 \Theta_{11}^0 (x, x - \beta)
\right\}
\nonumber\\
&-&\int d\beta' D(\beta , \beta')
\Biggl\{
2
\left[
\frac{ \beta }{(x - \beta)}
 \Theta_{11}^0 (x, x - \beta)
\right]_+
+ \frac{x}{x - \beta}
\Theta_{111}^0 (x , x - \beta, x - \beta + \beta') \nonumber\\
&+&\delta (\beta - x) \int d \beta '' \frac{\beta}{\beta''}
\Theta_{111}^0 (\beta'',\beta'' - \beta, \beta'' - \beta')
+
2 \Theta_{11}^0 (x, x - \beta)
\Biggr\}
\Biggr),\\
&&\nonumber\\
\dot D (x , x')&=&
-\frac{\alpha}{2\pi}
\Biggl\{
\left[
\frac{x'}{x}e(x) - M (x)
\right]
\Theta_{11}^0 (x', x' - x)
-
\left[
\frac{x}{x'}e(x') - M (x')
\right]
\Theta_{11}^0 (x, x - x')
\nonumber\\
&+&\int d\beta'
\Biggl(
D(x, \beta')\frac{(\beta' - x + x')}{(x - x')}
\Theta_{111}^0 (x', x' - x , x' - x + \beta')\nonumber\\
&+&\frac{x'}{x' - \beta'}
[D (x - x' + \beta', \beta') - D (x, x')]
\Theta_{11}^0 (x' , x' - \beta')
\Biggr) \nonumber\\
&+&\int d\beta
\Biggl(
D(\beta , x')\frac{(\beta - x' + x)}{(x' - x)}
\Theta_{111}^0 (x, x - x' , x - x' + \beta)\nonumber\\
&+&\frac{x}{x - \beta}
[D (\beta , x' - x + \beta) - D (x, x')]
\Theta_{11}^0 (x , x - \beta)
\Biggr) - \frac{3}{2}D(x,x')
\Biggr\},
\end{eqnarray}
where the dot denotes the derivative with respect to the UV cutoff
$\hspace{0.1cm}\dot{}=\Lambda^2{\partial}/{\partial\Lambda^2}$ and
the plus-prescription is defined by the equation
\begin{equation}
\label{plus}
\left[ \frac{\beta}{( x - \beta )}
\Theta_{11}^0 (x, x - \beta) \right]_+
=
\frac{ \beta }{( x - \beta )}
\Theta_{11}^0 (x, x - \beta)
-
\delta (\beta - x)
\int d \beta''
\frac{\beta}{ (\beta'' - \beta ) }
\Theta_{11}^0 ( \beta'', \beta'' - \beta).
\end{equation}
We have used also the equation
\begin{equation}
{\rm PV}\int d\beta \frac{x}{(x - \beta)}
\left[
\Theta^0_{11} ( \beta , \beta - x)
+\Theta^0_{11}(x, x - \beta)
\right]=0.
\end{equation}

By exploiting the relation provided by the equation of motion, we
can easily verify that the RG equations thus constructed are indeed
correct and the renormalization program can be reduced to the study
of logarithmic divergences of the three-parton $Z(x,x')$ and quark
mass $M(x)$ correlators in perturbation theory.

\subsection{QCD evolution of the twist-3 distributions.}

For the non-abelian gauge theory the equality of the renormalization
constants $Z_{\rm g} = Z_3$ implied above no longer holds; so we
should account for the renormalization of the gluon wave-function as
well as for the renormalization of charge explicitly. For these
purposes, to complete the renormalization program outlined in the
preceding section \ref{renormalization}, we evaluate the gluon
propagator to the same accuracy. The result can be written in the
compact form
\begin{equation}
D_{\mu\nu}(k) = \left( 1+\Pi^{tr}(k) \right)
U_{\mu\rho}(k)\frac{d_{\rho\sigma}(k)}{k^2+i0}
U_{\sigma\nu}(k),
\end{equation}
where
\begin{equation}
U_{\mu\nu}(k)=g_{\mu\nu}-\frac{1}{2}\Pi^{add}(k)
\frac{k_\mu n_\nu + k_\nu n_\mu}{k_+}
\end{equation}
and
\begin{eqnarray}
&&\Pi^{tr}(k)=2 \frac{\alpha}{4\pi} \ln \Lambda^2
\left\{
C_A\int dz \frac{[z^2-z\zeta +\zeta^2]^2}{z(z-\zeta)\zeta^2}
\Theta_{11}^0 (z,z-\zeta)
-\frac{N_f}{3}
\right\},\nonumber\\
&&\Pi^{add}(k)= \frac{\alpha}{4\pi} \ln \Lambda^2
C_A\int dz
\frac{[5z\zeta^2(z-\zeta) + 6z^2(z-\zeta)^2
+ 2\zeta^4]}{z(z-\zeta)\zeta^2}
\Theta_{11}^0 (z,z-\zeta)
\end{eqnarray}
are the transverse and longitudinal pieces of polarization operator.
The renormalized charge is given by the well-known "asymptotic
freedom" formula
\begin{equation}
{\rm g}_0
={\rm g}\left[
1 + \frac{\alpha}{4\pi}\ln \Lambda^2
\left(
\frac{N_f}{3} - \frac{11}{6}C_A
\right)
\right].
\end{equation}

Now we are in a position to adduce the RG equations for the real QCD
case. Just giving the final result (without intermediate steps) for
the chiral-even distributions we then address in grater detail to
the chiral-odd evolution.

\subsubsection{Evolution of the chiral-even distributions.}
\label{g_2-evolution}

Let us begin with the transversely polarized structure function $g_2$.
In the first papers \cite{first-g_2} on the logarithmic $Q^2$-variation
of the moments of $\widetilde g_2$ a simple evolution has been derived:
\begin{equation}
\label{DGLAP-g_2}
\int_{0}^{1} dx x^n \widetilde g_2 (z, Q)
= \left( \frac{\alpha(Q)}{\alpha(Q_0)} \right)
^{\gamma^g_n/\beta_0}
\int_{0}^{1} dx x^n \widetilde g_2 (z, Q_0),
\end{equation}
and the anomalous dimensions have been found by calculating the
radiative corrections to the operator
\begin{equation}
\label{Otw-3}
{\cal O}^3_{\sigma \mu_1 \mu_2 \dots \mu_n}
=i^n\Asym_{\sigma \mu_1}\Sym_{\mu_1\mu_2\dots\mu_n}
\bar\psi \gamma_\sigma \gamma_5
D_{\mu_1}D_{\mu_2}\dots D_{\mu_n} \psi
\end{equation}
over the free quark states. As we have seen above, the conclusion of
these works is erroneous since (\ref{Otw-3}) mixes with other
operators of the same twist and quantum numbers due to
renormalization. The most significant departure from the naive
expectation is that the number of operators involved in the RG
evolution turns out to be increasing with the moment \cite{lip84}, and
as a result it is impossible to write the equation of the DGLAP type
which manages corresponding scale dependence for the $\widetilde g_2$.
Moreover, the eigenvalues of the anomalous dimension matrix are not
known analytically. However, an way out has been found \cite{ali91} in
the two important limits: $N_c \to \infty$ and $x \to 1$, where
evolution reduces to the DGLAP equation (\ref{DGLAP-g_2}), though
the anomalous dimension turns out to be different from that found
in \cite{first-g_2}. The combining use of these asymptotics does
provide an excellent approximation \cite{bel97} to the complete result.

To simplify the discussion, we neglect the quark-mass operator. The
nonlocal string operators ${^\pm\!{\cal S}_\rho}$ introduced
previously are related to each other by charge conjugation (so are
the corresponding kernels which govern their evolution) and do not mix
in the course of renormalization; therefore, we give only the
result\footnote{This discussion follows Ref. \cite{mul96}}
for ${^-\!{\cal S}_\sigma}$
\begin{eqnarray}
{^-\!\dot{\cal S}_\sigma}(\mu, \lambda) &=&
\frac{\alpha}{2\pi}
\int_0^1 dy \int_0^{\bar y} dz
\biggl\{
C_A \biggl[
\left[ [N(y,z)]_+ - {7\over 4} \delta(\bar{y})\delta(z) \right]
{^-\!{\cal S}_\sigma}(\mu y, \lambda - \mu z) \nonumber\\
&+& \left[ 2 \bar z + [N(y,z)]_+
- {7\over 4} \delta(\bar{y})\delta(z) \right]
{^-\!{\cal S}_\sigma}(\mu - \lambda z, \lambda y)
\biggr] \nonumber\\
&+& \left( C_F-\frac{C_A}{2} \right)
\biggl[ y\,\delta (z)
{^-\!{\cal S}_\sigma}(-\mu y, \lambda - \mu y)
- 2z {^-\!{\cal S}_\sigma}(\mu - \lambda \bar z, - \lambda y) \nonumber\\
&+& [K(y,z)]_+ {^-\!{\cal S}_\sigma}
(\mu \bar z + \lambda z, \lambda \bar y + \mu y) \biggr]
\biggr\}.
\end{eqnarray}
To condense the notation we have used the dot as short-hand for the
logarithmic derivative with respect to the renormalization scale
$\hspace{0.1cm}\dot{}=\mu_R^2\,{\partial}/{\partial\mu_R^2}$ and
${^-\!{\cal S}^\rho}(\mu,\lambda)
= {^-\!{\cal S}^\rho}(\mu, 0, \lambda)$. The standard
plus-prescription fulfilling is
\begin{eqnarray}
\label{N-funct}
&&\hspace{-1cm}[N(y,z)]_+ = N(y,z)
- \delta(\bar y) \delta(z)
\int_{0}^{1} dy' \int_{0}^{\bar y'} dz' N(y',z'), \hspace{0.5cm}
N(y,z) = \delta (\bar y-z) \frac{y^2}{\bar y}
+ \delta (z) \frac{y}{\bar y},\nonumber\\
&&\hspace{-1cm}[K(y,z)]_+ = K(y,z)
- \delta(y) \delta(z)
\int_{0}^{1} dy' \int_{0}^{\bar y'} dz' K(y',z'), \hspace{0.5cm}
K(y,z) = 1 + \delta (y) \frac{\bar z}{z} + \delta (z) \frac{\bar y}{y}.
\nonumber\\
\end{eqnarray}

To proceed further we construct a C-even operator from the
Shuryak-Vainshtein ones
\begin{equation}
{\cal Y}_\sigma(\lambda, \mu) =
{^+\!{\cal S}^\rho}(\lambda, \mu) + {^-\!{\cal S}^\rho}(\mu, \lambda)
\end{equation}
and define a new distribution function as Fourier transform with
respect to the variable $\lambda$ only, so that
\begin{equation}
{\scriptstyle\cal Y} (x,u) S^\perp_\sigma =
\frac{1}{2}
\int \frac{d\lambda}{2\pi} e^{i \lambda x }
\langle h | {\cal Y}_\sigma (\bar u \lambda, - u \lambda)
+ (\lambda \to - \lambda) | h \rangle
\end{equation}
depends on the effective momentum fraction $x$ and the gluon
position on the light cone $u$.

Skipping the details (to which we address in the following discussion
of the chiral-odd distributions), we just note that the genuine twist-3
part of $g_2$ which can be expressed via the integral of the
three-parton correlator
\begin{equation}
\widetilde g_2 (x) = - \bar g_2(x)
+ \int_{x}^{1} \frac{dy}{y} \bar g_2(x)
\end{equation}
with
\begin{equation}
x \bar g_2(x) = - \frac{d}{dx} \int_{0}^{1} du u
{\scriptstyle\cal Y} (x,u),
\end{equation}
satisfies the DGLAP evolution equation in the large-$N_c$ limit
({\it i.e.} neglecting the terms ${\cal O}(1/N_c^2)$)
\begin{equation}
\dot{[x \bar g_2(x)]} = \frac{\alpha}{4\pi}
\int_{x}^{1} \frac{dy}{y} P_{gg}\left( \frac{x}{y} \right)
[y \bar g_2(y)],
\end{equation}
with the splitting function \cite{ali91,mul96}
\begin{equation}
\label{split-N_c-g_2}
P_{gg}(y) = N_c \left\{
\left[ \frac{2}{\bar y} \right]_+ - 2 - y
+ \frac{1}{2} \delta (\bar y)
\right\}.
\end{equation}
In the $x \to 1$ region the evolution equation remains of the
DGLAP type even for the $\frac{1}{N_c}$-suppressed contribution.
The combining use of the Eq. (\ref{split-N_c-g_2}) and an additional
piece
\begin{equation}
\Delta P_{gg} (y) = - \frac{1}{N_c}
\left\{  \left[ \frac{2}{\bar y} \right]_+
+ \frac{3}{2} \delta (\bar y)\right\}.
\end{equation}
for ${\cal O}(1/N_c)$-terms with $x-1 \ll 1$ yields a good
approximation for the exact evolution. Ob\-vi\-ous\-ly, the function
$\widetilde g_2(x)$ obeys the same evolution equation as
$\bar g_2$. Actually, the splitting function (\ref{split-N_c-g_2})
has been exploited in Ref. \cite{strat93} to rescale the bag model
predictions to values of $Q^2$ of the real experiment. It has been
shown there that $g_2^{WW}(x)$ and $\widetilde g_2(x)$ enter 
into $g_2(x)$ on equal footing at the model scale $\mu^2_{\rm bag}$. 
Moreover, it has been figured out that this situation does not changed 
at higher $Q^2$ and $\widetilde g_2(x)$ remains an important 
ingredient of $g_2(x)$.

\subsubsection{Evolution of the chiral-odd
distributions\protect\footnote{The following discussion mimics
the corresponding sections from Ref. \protect\cite{bel97}}.}

Turning to the case of the chiral-odd distributions, we have to note
that in the leading logarithmic approximation the evolution equations
that govern the $Q^2$-dependence of the three-particle correlation
functions are the same, discarding the mixing with the quark mass
operator. Therefore, we omit the "tilde" sign in what follows.

In the light-cone fraction representation we get for the
correlation function $D(x,x')$
\begin{eqnarray}
\label{eveqz}
&&\hspace{-1cm}\dot D(x , x')=
-\frac{\alpha}{2\pi}
\Biggl\{
-C_F \frac{(x - x')}{x x'}
\left[
x' M (x) \Theta_{11}^0 (x', x' - x)
\pm
x M (x') \Theta_{11}^0 (x, x - x')
\right]
\nonumber\\
&&\hspace{-1cm}+
\int d\beta
\Biggl(
C_F  D (\beta , x')\frac{x}{x'}
\Theta_{11}^0 (x, x - x')
+\frac{C_A}{2}
\biggl(
[D (\beta , x') - D (x , x')]\frac{x}{(x - \beta)}
\Theta_{11}^0 (x, x - \beta) \nonumber\\
&&\hspace{-1cm}+\ [D (\beta + x', x') - D (x , x')]
\frac{(x - x')}{(x - x' - \beta)}
\Theta_{11}^0 (x - x', x - x' - \beta) \nonumber\\
&&\hspace{-1cm}+\ \frac{(\beta + x - x')}{x'}
\biggl(
D (\beta , x')\frac{x}{(x' - x)}
\Theta_{11}^0 (x , x - \beta)
+
D (\beta + x', x')\Theta_{11}^0
(x - x' , x - x' - \beta)
\biggr)
\biggr)\nonumber\\
&&\hspace{-1cm}+
\left(
C_F - \frac{C_A}{2}
\right)
\biggl(
D(\beta , x')\frac{(\beta + x - x')}{(x' - x)}
\Theta_{111}^0 (x, x - x' , x - x' + \beta)\nonumber\\
&&\hspace{-1cm}+\
[D (\beta , x' - x + \beta) - D (x, x')]\frac{x}{x - \beta}
\Theta_{11}^0 (x , x - \beta)
\biggr)
\Biggr)\nonumber\\
&&\hspace{-1cm}+
\int d\beta'
\Biggl(
C_F  D (x , \beta')\frac{x'}{x}
\Theta_{11}^0 (x', x' - x)
+\frac{C_A}{2}
\biggl(
[D (x , \beta') - D (x , x')]\frac{x'}{(x' - \beta')}
\Theta_{11}^0 (x', x' - \beta') \nonumber\\
&&\hspace{-1cm}+\ [D (x , \beta' + x) - D (x , x')]
\frac{(x' - x)}{(x' - x - \beta')}
\Theta_{11}^0 (x' - x, x' - x - \beta') \nonumber\\
&&\hspace{-1cm}+\ \frac{(\beta' + x' - x)}{x}
\biggl(
D (x , \beta')\frac{x'}{(x - x')}
\Theta_{11}^0 (x' , x' - \beta')
+
D (x , \beta' + x)\Theta_{11}^0
(x' - x , x' - x - \beta')
\biggr)
\biggr)\nonumber\\
&&\hspace{-1cm}+
\left(
C_F - \frac{C_A}{2}
\right)
\biggl(
D( x ,\beta' )\frac{(\beta' + x' - x)}{(x' - x)}
\Theta_{111}^0 (x', x' - x , x' - x + \beta')\nonumber\\
&&\hspace{-1cm}+\
[D ( x - x' + \beta' , \beta') - D (x, x')]\frac{x'}{x' - \beta'}
\Theta_{11}^0 (x' , x' - \beta')
\biggr)
\Biggr) - \frac{3}{2}C_F D(x,x')
\Biggr\},
\end{eqnarray}
and for the mass-dependent correlation function we have
\begin{equation}
\label{mass}
\dot M(x)
= - C_F \frac{\alpha}{2\pi}\int d \beta M(\beta)
\left\{
2
\left[
\frac{ \beta }{( x - \beta )}\Theta_{11}^0 (x, x - \beta)
\right]_+
+
\frac{\beta + x}{\beta} \Theta_{11}^0 (x, x-\beta)
\right\},
\end{equation}
where we have used the standard plus-prescription fulfilling
$\int dx [...]_+ = 0$ (for a definition see Eq. (\ref{plus})).
Throughout the paper the plus and minus signs in the mass-operator
term correspond to the functions $D$ (for $e$) and $\widetilde D$
(for $h_L$), respectively.

For the string operators we obtain the following compact RG equation:
\begin{eqnarray}
\label{evolequaZ}
\dot{\cal Z}(\lambda , \mu) &=&
\frac{\alpha}{2 \pi}
\int_{0}^{1} dy \int_{0}^{\bar y} dz
\biggl\{
C_F {\bar y}^2 \delta (z)
\left[
{\cal M}^1 (\lambda - \mu y)
\pm
{\cal M}^1 (\lambda y - \mu)
\right] \nonumber \\
&+&\frac{C_A}{2}
\left[  2\bar z + [N(y,z)]_+ -\frac{7}{4} \delta(\bar y) \delta(z)
\right]
\left[ {\cal Z}(\lambda y, \mu - \lambda z)
+ {\cal Z}(\lambda - \mu z, \mu y) \right] \nonumber\\
&+& \left( C_F - \frac{C_A}{2} \right)
\biggl[
\left[ [L(y,z)]_+ - \frac{1}{2} \delta(y) \delta(z) \right]
{\cal Z}(\lambda \bar z + \mu z , \mu \bar y + \lambda y)
\nonumber\\
&-& 2z \left[
{\cal Z}(-\lambda y , \mu - \lambda \bar z)
+ {\cal Z}(\lambda - \mu \bar z , -\mu y)
\right]
\biggr]
\biggr\},
\end{eqnarray}
with the function $N(y, z)$ given by Eq. (\ref{N-funct}) and
\begin{equation}
L(y,z) = \delta (y) \frac{\bar z}{z} + \delta (z) \frac{\bar y}{y}.
\end{equation}
The equations written so far should be supplemented by the following
\begin{eqnarray}
\label{mass-NL}
\dot{\cal M}^1 (\lambda) &=& \frac{\alpha}{2\pi}C_F
\int_{0}^{1}dy
\left\{ \left[ \frac{2}{\bar y} \right]_+ - 2 - y - y^2 \right\}
{\cal M}^1 (\lambda y) ,\\
\dot h_1 (\lambda) &=& \frac{\alpha}{2\pi}C_F
\int_{0}^{1}dy
\left\{ \left[ \frac{2}{\bar y} \right]_+
- 2 + \frac{3}{2}\delta (\bar y) \right\}
h_1 (\lambda y).
\end{eqnarray}
The last one, when transformed to the momentum space using the
formulae of the next section, coincides with the result obtained in
Ref. \cite{art90}.

\subsection{Local anomalous dimensions.}

Now we are able to pass from the evolution equations for correlators
to the equations for their moments and find, in this way, the
anomalous dimension matrix for local twist-3 quark-gluon operators.

We define the moments in following way:
\begin{eqnarray}
\label{mom-2-3}
F_n &=& \int dx x^n F(x)\hspace{0.5cm}\mbox{for any two-particle
correlator,}\nonumber\\
\label{defmom}
Z_n^l &=& \int dx dx' x^{n-l} x'^{l-1} Z (x,x').
\end{eqnarray}
In the language of operator product expansion these equalities
specify the expansion of nonlocal string operators in towers of
the local ones, namely
\begin{eqnarray}
{\cal Z}_n^l &=&
i^{n-1}(-1)^{l-1}
\frac{\partial^{l-1}}{\partial\mu^{l-1}}
\frac{\partial^{n-l}}{\partial\lambda^{n-l}}
\left.{\cal Z}(\lambda , \mu)\right|_{\lambda=\mu=0} \nonumber\\
&&\hspace{5cm}= \frac{1}{2} \bar \psi (0) (iD_+)^{l-1}
{\rm g} G_{+ \rho} (0) \sigma^\perp_{\rho +}
\!\left( \!
\begin{array}{c}
I \\
\gamma_5
\end{array}
\!\right)
(iD_+)^{n-l}\psi (0),\nonumber\\
{\cal M}_n &=&
i^n \frac{\partial^n}{\partial\lambda^n}\,
\left.{\cal M}(\lambda)\right|_{\lambda=0}
=\frac{m}{2}\bar \psi (0)
\gamma_+
\!\left( \!
\begin{array}{c}
I \\
\gamma_5
\end{array}
\!\right)
(iD_+ )^n \psi (0).
\end{eqnarray}
The inverse transformations to the nonlocal representation are
given by
\begin{eqnarray}
{\cal Z}(\lambda,\mu)
=\sum_{n=0,\,m=0}^{\infty} (-i)^{n+m}
(-1)^m\frac{\mu^m}{m!}\frac{\lambda^n}{n!}\,
{\cal Z}_{n+m+1}^{m+1},\qquad
{\cal M}(\lambda)
=\sum_{n=0}^{\infty} (-i)^{n} {\lambda^n\over n!}\,{\cal M}_n.
\end{eqnarray}

Now it is a simple task to derive the algebraic equations
for the mixing of local operators under the change of the
renormalization scale from the evolution equations
(\ref{eveqz})-(\ref{mass-NL}). They are
\begin{eqnarray}
\label{comparison-1}
\dot M_n &=& \frac{\alpha}{2\pi}
{_{\scriptscriptstyle MM}\gamma}^n M_n,\\
\dot Z^l_n &=& \frac{\alpha}{2\pi}
\left\{
\left[ {_{\scriptscriptstyle ZM}\gamma}^n_{n-l+1}
\pm {_{\scriptscriptstyle ZM}\gamma}^n_l \right] M_n
+\sum_{k=1}^{n}{_{\scriptscriptstyle ZZ}\gamma}^n_{lk} Z_n^k
\right\},
\end{eqnarray}
where the anomalous dimensions are given by the compact expressions
\begin{eqnarray}
\label{comparison-2}
{_{\scriptscriptstyle MM}\gamma}^n
&=& - C_F \left( S_n + S_{n+2} \right), \\
{_{\scriptscriptstyle ZM}\gamma}^n_l
&=& \frac{2C_F}{l(l+1)(l+2)}, \\
\label{anomdimZ}
{_{\scriptscriptstyle ZZ}\gamma}^n_{lk}
&=& \frac{3}{4}C_F \delta (l - k)
+\frac{C_A}{2}
\left\{
\theta (l-k-1)\frac{(k+1)(k+2)}{(l-k)(l+1)(l+2)}
-\delta (l-k)\left[ S_{k-1} + S_{k+2} \right]
\right\}\nonumber\\
&+& \left( C_F - \frac{C_A}{2} \right)
\biggl\{
\theta (l-k-1)
\left[
\frac{2(-1)^k C_l^k}{l(l+1)(l+2)}
+\frac{(-1)^{l-k}}{(l-k)}\frac{C_n^{k-1}}{C_n^{l-1}}
\right] \nonumber\\
&+& \delta (l-k)\left[
\frac{2(-1)^k}{k(k+1)(k+2)} -S_k
\right]
\biggr\}
+ {k \rightarrow n-k+1 \choose l \rightarrow n-l+1}.
\end{eqnarray}
Here, we have used the following step functions:
\begin{equation}
\label{Kron}
\theta (i-j)
= \left\{
\begin{array}{c}
1, \ i \geq j \\
0, \ i<j
\end{array}
\right. ,
\hspace{1cm}
\delta (i-j)
= \left\{
\begin{array}{c}
1, \ i = j \\
0, \ i \neq j
\end{array}
\right. ,
\end{equation}
as well as the convention
$ S_n = \sum_{k=1}^{n} \frac{1}{k}$ and the binomial coefficients
$C_n^m = \frac{n!}{m!(n-m)!}$.
The results of this section have been independently derived in Ref.
\cite{koi95} using the standard approach based on the local operator
product expansion.

\subsection{Relating the evolution kernels in the light-cone
position and light-cone fraction re\-pre\-sen\-ta\-tions.}

Until recently the relation between different formulation of
the evolution equations in the light-cone fraction \cite{lip84,lip85}
and light-cone position \cite{bb88,mul94} representations has been
obscure and has been thought to be difficult to realize \cite{folk}.
However, having at hand the evolution equations in different
representations for the same quantities, we are able to fill this
gap and relate the kernels in both the cases \cite{bel97}. Such a bridge
can be easily established using the Fourier transformation for the
parton distribution functions given by Eqs. (\ref{Fourier-2}) and
(\ref{Fourier-3}).

To start with, we address ourselves to a simpler case of the
two-particle correlation functions ${\cal F}$. The evolution
equation in the light-cone position space is of the following
generic form:
\begin{equation}
\dot{\cal F} (\lambda)
= \int_{0}^{1} dy {\cal K} (y) {\cal F} (\lambda y),
\end{equation}
where ${\cal K} (y)$ is the corresponding evolution kernel. By exploiting
the definitions (\ref{Fourier-2}) we can recast the Fourier transform
on the language of two-particle evolution kernels. In this way, we find
the direct transformation
\begin{equation}
K(x,\beta) = \int_{0}^{1} dy {\cal K}(y) \delta (x - y\beta).
\end{equation}
With the help of the general formula
\begin{equation}
\label{theta}
\int_{0}^{1} dy f(y) \delta (x - y\beta)
= f\left( \frac{x}{\beta} \right)
\Theta^0_{11} (x, x - \beta).
\end{equation}
we observe that the RG equations for two-parton correlators
derived in the previous section are the same indeed. The inverse
transformation can be done
\begin{equation}
\int \frac{dxd\beta}{2\pi} e^{-i\lambda x + i\mu \beta}
K(x, \beta) = \int_{0}^{1}dy {\cal K}(y) \delta (\mu - y \lambda)
\end{equation}
with the following result
\begin{equation}
\int \frac{dxd\beta}{2\pi} f\left( \frac{x}{\beta} \right)
\Theta^0_{11} (x, x - \beta) e^{-i\lambda x + i\mu \beta}
= \int_{0}^{1} dy f(y) \delta (\mu - y \lambda).
\end{equation}

The transformation for three-particle correlators is a little bit
more involved. The general form of the evolution equation for the
light-cone string operator ${\cal Z}(\lambda,\mu)$ reads
\begin{equation}
\dot{\cal Z}(\lambda , \mu)
=\int_{0}^{1}dy\int_{0}^{\bar y}dz {\cal K}(y,z)
{\cal Z} (\eta_{11}\lambda + \eta_{12}\mu,
\eta_{21}\lambda + \eta_{22}\mu),
\end{equation}
where $\eta_{ij}$ are linear functions of the variables $y$, $z$.
In the momentum fraction representation the evolution equation
looks like
\begin{equation}
\dot Z(x,x') = \int d\beta d\beta'
K (x,x',\beta ,\beta') Z(\beta,\beta').
\end{equation}
Specifying the particular form of the functions $\eta_{ij}$,
we display below an example for the conversion supplied with a
general formula suitable for all practical cases of interest.

For the $C_A/2$ part of the evolution equation, the Fourier
transformation gives
\begin{equation}
K(x,x',\beta,\beta') = \delta (\beta' - x')
\int_{0}^{1}dy\int_{0}^{\bar y}dz
{\cal K}(y,z)\delta (x - x'z - \beta y).
\end{equation}
The particular contribution is
\begin{eqnarray}
&&\hspace{-0.5cm}{\cal K} (y,z) = z
\ \stackrel{FT}{\rightarrow} \ K (x,x',\beta,\beta') =
\delta (\beta' - x')
\biggl\{
\frac{x - \beta}{x'} \Xi_1 (x, x - x', x - \beta)\nonumber\\
&&\hspace{7.8cm}+ \frac{\beta}{x'} \Xi_2 (x, x - x', x - \beta)
\biggr\}.
\end{eqnarray}
Here, we have used (\ref{theta}) and the following general result:
\begin{eqnarray}
\label{Xi}
&&\Xi_n (x, x-x', x-\beta) \equiv
\int_{0}^{1} dy y^n \Theta^0_{11}
((x-\beta)+y\beta,(x-\beta)-y(x'-\beta))\\
&&=\frac{1}{n}\left[ 1
- \left( \frac{\beta - x}{\beta - x'} \right)^n \right]
\Theta^0_{11} (x, x-x')
+\frac{1}{n}\frac{\beta}{x'}
\left[ \left( \frac{\beta - x}{\beta - x'} \right)^n
- \left( \frac{\beta - x}{\beta} \right)^n
\right]
\Theta^0_{11} (x, x-\beta).\nonumber
\end{eqnarray}

The complete list of transformations can be found in Ref.
\cite{bel97}. Using these results, we can easily verify that the
evolution equations given by Eqs. (\ref{eveqz}) and (\ref{evolequaZ})
agree with each other. It should be noted that it is sufficient to
have at hand Eqs. (\ref{theta}) and (\ref{Xi}) to perform the
conversion from one representation to another.

\subsection{Generalized DGLAP equations for the three-parton
cor\-re\-la\-tors.}

As we have seen above, the evolution equations in the momentum
space (\ref{eveqz}) turn out to be very complicated, while
Eq. (\ref{evolequaZ}) being compact is not suitable for an analysis
since only its Fourier transform is related to the physical
observables. Therefore, to obtain a simple and manageable equation
which can be attempted to be diagonalized, we are forced to proceed
further in the same line as suggested in section \ref{g_2-evolution}
for the chiral-even structure function. For this purpose we define
a new function, Fourier-transformed with respect to the $\lambda$
variable only:
\begin{eqnarray}
\label{def3part}
{\scriptstyle\cal Z}(x,u)
= \frac{1}{2} \int \frac{d\lambda}{2\pi}
e^{i\lambda x }
\langle h | {\cal Z}(\bar u \lambda, -u \lambda)
\pm ( u \to \bar u ) |h \rangle ,
\end{eqnarray}
which is even under charge conjugation and depends on the variables
$x$ and $u$. The latter has the meaning of the relative position of
the gluon field on the light cone. For $0\le u \le 1$ the gluon
field lies between the two quark fields. Because of the support
property $|x|\leq {\rm max}(1,|2u-1|)$, the variable $x$ is then
restricted to $|x|\leq 1$ and can be interpreted as an effective
momentum fraction.

The evolution equation for ${\scriptstyle\cal Z}(x,u)$
can be derived in a straightforward way from the RG equation
(\ref{evolequaZ}) for the nonlocal string operator ${\cal Z}$.
It can be presented in the form of a generalized DGLAP-type
equation in the mixed representation\footnote{This particular
representation of the evolution equation for three-particles
distributions has first been given in the second paper of
Ref. \cite{mul96}.}:
\begin{eqnarray}
\label{EvoEqux&u}
\dot {\scriptstyle\cal Z}\left( x,u \right)
&=&\frac{\alpha}{2\pi}
\int \frac{dy}{y} \int dv
\Bigg\{
P_{{\scriptscriptstyle\cal Z Z}}(y,u,v)
{\scriptstyle\cal Z} \left( \frac{x}{y},v \right)
+
P_{{\scriptscriptstyle\cal Z} m }(y,u,v)
m \left( \frac{x}{v} \right)
\Bigg\}, \\
\label{EvoEqux&u-mass}
\dot m(x)
&=&\frac{\alpha}{2\pi}
\int \frac{dy}{y} P_{mm}(y)
m \left( \frac{x}{y} \right).
\end{eqnarray}
Here, $m(x)=xM(x)$ and the integration region is determined by both
the support of ${\scriptstyle\cal Z}(x,u)$ and the kernels
\begin{eqnarray}
&&P_{\scriptscriptstyle\cal Z Z}(x,u,v) \nonumber\\
&&=\left( C_F - \frac{C_A}{2} \right)
\Bigg[ {\mit\Theta}_1(x,u,v) [L(x,u,v)]_+
-{\mit\Theta}_2(x,u,v) M(x,u,v)
-{1\over 4} \delta(u-v) \delta(\bar x)\Bigg] \nonumber\\
&&+\frac{C_A}{2}{\mit\Theta}_3(x,u,v)
\left[ M(x,u,v) + [N(x,u,v)]_+
-\frac{7}{4} \delta(u-v)\delta(\bar x)\right]
+ {u \rightarrow \bar u \choose v \rightarrow \bar v},\\
&&\nonumber\\
&&P_{{\scriptscriptstyle\cal Z} m}(x,u,v)
=C_F {\bar x}^2\theta(x)\theta(\bar x) \frac{x}{v}
\left[ \delta(v-\bar u - x u) \pm \delta(v-u-x\bar u ) \right],\\
&&\nonumber\\
\label{Pmm}
&&P_{mm}(x)
=C_F \left[
\left[\frac{2}{\bar x}\right]_+ - 2 - x - x^2
\right],
\end{eqnarray}
where the auxiliary functions are defined by:
\begin{eqnarray}
{\mit\Theta}_1(x,u,v)&=&
\theta(x) \theta(u - x v) \theta(\bar u - x \bar v ), \nonumber\\
{\mit\Theta}_2(x,u,v)&=&
\theta\left( -\frac{x\bar v}{\bar u} \right)
\theta\left( \frac{1- x v }{\bar u} \right)
\theta\left( \frac{x - u}{\bar u} \right), \nonumber\\
{\mit\Theta}_3(x,u,v)&=&
\theta\left( \frac{\bar x}{\bar u} \right)
\theta\left( \frac{x \bar v}{\bar u} \right)
\theta\left( \frac{x v - u}{\bar u} \right),
\nonumber\\
L(x,u,v) &=& \frac{u^2}{v (v-u)}\delta(u - x v) ,
\nonumber\\
M(x,u,v) &=& \frac{2 x (1 - x v)}{{\bar u}^3},
\nonumber\\
N(x,u,v) &=& \frac{\bar v \epsilon (\bar u)}{\bar u (v-u)}
\left[ \frac{\bar v}{\bar u} \delta(\bar x)
+ \frac{u^2}{v} \delta(u - x v) \right].
\end{eqnarray}
The plus-prescription for the arbitrary function $A$ is defined by
the equation
\begin{eqnarray}
&&{\mit\Theta}_i(x,u,v) [A(x,u,v)]_+ \nonumber\\
&&= {\mit\Theta}_i(x,u,v) A(x,u,v)- \delta(\bar x) \delta(u-v)
\int_0^1 dx'\int dv'\, {\mit\Theta}_i(x',u,v') A(x',u,v').
\end{eqnarray}
Note that, due to the evolution, the variable $u$ is no longer
restricted to the region $0\le u \le 1$.

\subsection{Eigenvalues and eigenfunctions.}

Going further, for simplicity, we restrict ourselves to the
homogeneous case, {\it i.e.} we discard the quark-mass operator,
which is certainly a justified assumption for the light $u$- and
$d$-quark species. The diagonalization of the evolution equation
(\ref{EvoEqux&u}) can be achieved by introducing the Mellin
transforms
\begin{equation}
{\scriptstyle\cal Z}^n(u) = \int dx x^{n-1} {\scriptstyle\cal Z}(x,u),
\end{equation}
where $n$ is the complex angular momentum. Operators with different $n$
do not mix with each other and satisfy the following equation:
\begin{eqnarray}
\label{Mellin-Z}
\dot{\scriptstyle\cal Z}^n(u) &=& \frac{\alpha}{2\pi}
\int dv
P^n_{{\scriptscriptstyle\cal Z Z}} (u,v)
{\scriptstyle\cal Z}^n (v).
\end{eqnarray}
With the kernel given by
\begin{eqnarray}
\label{DefPZZ}
&&P^n_{{\scriptscriptstyle\cal Z Z}}(u,v)
= \left( C_F - \frac{C_A}{2} \right)
\Bigg[ {\mit\Theta}_1(u,v) [L^n(u,v)]_+
- {\mit\Theta}_2 (u,v) M^n_1(u,v)
- \frac{1}{4} \delta(u-v)\Bigg] \nonumber\\
&&+ \frac{C_A}{2}{\mit\Theta}_3(u,v)
\left[ M^n_2(u,v) + [N^n(u,v)]_+ - \frac{7}{4} \delta(u-v) \right]
+ {u \rightarrow \bar{u} \choose v\rightarrow \bar{v}},
\end{eqnarray}
where the auxiliary functions read
\begin{equation}
{\mit\Theta}_1(u,v) = \theta(v-u), \quad
{\mit\Theta}_2(u,v)= \theta(- \bar v) \theta(1 - v u), \quad
{\mit\Theta}_3(u,v)= \theta(\bar v)\theta(v-u).
\end{equation}
\begin{eqnarray}
L^n(u,v)
&=& \frac{\epsilon (v)}{v-u}
\left( \frac{u}{v} \right)^{n+1}, \nonumber\\
M_1^n(u,v)
&=& \frac{2}{{\bar u}^3}
\left\{
\frac{1}{n+1} \left[ \frac{1}{v^{n+1}} - u^{n+1} \right]
- \frac{v}{n+2} \left[ \frac{1}{v^{n+2}} -u^{n+2} \right]
\right\},\nonumber\\
M_2^n(u,v)
&=& \frac{2}{{\bar u}^3}
\left\{
\frac{1}{n+1} \left[ 1 - \left(\frac{u}{v} \right)^{n+1} \right]
- \frac{v}{n+2} \left[ 1 - \left(\frac{u}{v} \right)^{n+2} \right]
\right\}, \nonumber\\
N^n(u,v)
&=& \frac{\bar v \epsilon(\bar u)}{\bar u (v-u)}
\left\{ \frac{\bar v}{\bar u}
+ \epsilon (v) \left( \frac{u}{v} \right)^{n+1} \right\}.
\end{eqnarray}
The plus-prescription is defined as
\begin{equation}
{\mit\Theta}_i(u,v) [A^n(u,v)]_+ =
{\mit\Theta}_i (u,v)A^n(u,v)
-\delta(u-v) \int dv' {\mit\Theta}_i(u,v') A^n(u,v').
\end{equation}
Here, we can check our calculation once more since in multicolour
limit, exact Eq. (\ref{Mellin-Z}) is reduced to the approximated
equation of Ref. \cite{bbkt96}.

\begin{figure}[htb]
\vspace{2cm}
\mbox{
\hspace{3cm}
\begin{picture}(170,110)
\put(0,0)	{
\epsffile{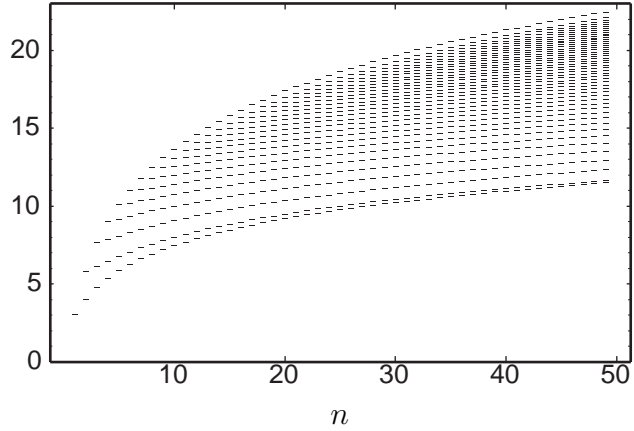}
		}
\put(51,72){$n$}
\end{picture}
}
\vspace{-7.6cm}
{\caption{\label{eigenvalues}
The spectrum of the eigenvalues $\lambda^n_l$ for the
evolution kernel $P^n_{\scriptscriptstyle\cal ZZ}$ defined in
(\protect\ref{DefPZZ}).
}}
\end{figure}

To obtain the solution of the evolution equation (\ref{Mellin-Z}), we
choose $n$ as a positive integer. In this case, as follows from the
definition (\ref{def3part}) of ${\scriptstyle\cal Z}^n(u)$ the
$n$-th moment is actually given by the following linear combination
of local operators ${\cal Z}_n^l$ (see Eq. (\ref{defmom})):
\begin{eqnarray}
\label{Z(u)bylocalO}
{\scriptstyle\cal Z}^n(u)
=\sum_{l=1}^{n} C_{n-1}^{l-1}\,
u^{n-l} {\bar u}^{l-1} {\cal Z}_n^l,
\end{eqnarray}
so that ${\scriptstyle\cal Z}^n(u)$ is a polynomial of degree $n-1$
in $u$. Thus, the kernel $P^n_{{\scriptscriptstyle\cal Z Z}}(u,v)$
possesses $n$ polynomial eigenfunctions $e^n_{l}(v)$
\begin{eqnarray}
\int dv\,P^n_{{\scriptscriptstyle\cal Z Z}}(u,v) e^n_{l}(v)
= -\lambda^n_{l} e^n_{l}(u),
\quad l=1,\dots,n,
\end{eqnarray}
where $-\lambda^n_{l} $ denotes the eigenvalues.
The solution we are
interested in is given in terms of the eigenvalues and
eigenfunctions of the anomalous-dimension matrix
$_{\scriptscriptstyle ZZ}\gamma_n^l$ of the local operators
${\cal Z}_n^l$. The eigenvalue problem we
have attacked has no analytical solution; however, the
diagonalization can be done numerically for a moderately
large orbital momentum $n$, e.g.\  $n\leq 100$, which is quite
sufficient for practical purposes.
These eigenfunctions
can be constructed by diagonalization
\begin{eqnarray}
\label{reltolocanomdim}
C_{n-1}^{k-1} \int dv\,P^n_{{\scriptscriptstyle\cal Z Z}}(u,v)
v^{n-k} \bar{v}^{k-1} &=&
\sum_{l=1}^{n} C_{n-1}^{l-1}
{_{\scriptscriptstyle ZZ}\gamma}^n_{lk} \, u^{n-l} \bar{u}^{l-1},
\end{eqnarray}
where the anomalous-dimension matrix
${_{\scriptscriptstyle ZZ}\gamma}^n_{lk}$ of the local operators
is given by Eq. (\ref{anomdimZ}). Actually, this
is a purely algebraic task and we find
\begin{eqnarray}
\label{eigenfunc}
e^n_{k}(u)
= \sum_{l=1}^{n} C_{n-1}^{l-1}\,u^{n-l} \bar{u}^{l-1} E^n_{lk},
\mbox{\ with\ }
\left\{ (E^n)^{-1} {_{\scriptscriptstyle ZZ}\gamma}^n E^{n}
\right\}_{kl} = -\lambda^n_{k}\delta (k - l),
\end{eqnarray}
where $\delta (k - l)$ is a Kronecker symbol defined by Eq.\
(\ref{Kron}). The spectrum of the eigenvalues $\lambda^n_l$ up to
$n=50$ is shown in Fig.~\ref{eigenvalues}. The solution for the moments
${\scriptstyle\cal Z} ^n(u)$ (in the massless case) is then
expressed in terms of the eigenfunctions and eigenvalues we have
found \cite{bel97}
\begin{eqnarray}
\label{solution}
{\scriptstyle\cal Z}^n\left(u,Q^2\right) =
\sum_{l=1}^{n} c^n_l\left(Q_0^2\right) e^n_{l}(u)
\exp{ \left\{
-\int_{Q_0^2}^{Q^2} \frac{dt}{t} \frac{\alpha(t)}{2\pi}
\lambda^n_{l}
\right\}}.
\end{eqnarray}
The coefficients $c^n_l\left(Q_0^2\right)$ at the
reference momentum squared $Q^2_0$ have to be determined from the
non-perturbative input.

\unitlength1mm

Obviously, in order to find the evolution of the higher $n$-moments
of the structure functions the whole information about the
relative size of the reduced matrix elements of the ($l=1,\dots,n$)
local quark-gluon operators is needed. At the moments, in the lack
of complete understanding of the yet unclear confinement mechanism
this problem is not accessible by nonperturbative methods presently
available. However, we should note that in the nearest future it
will not be possible to distinguish experimentally between the terms
with different anomalous dimensions even for the transverse
structure function $g_2(x)$ to say nothing of $h_L(x)$ (not to
mention $e(x)$).

Eq. (\ref{solution}) can be rewritten directly for the structure
functions entering into the physical cross sections. To this end, we
have to dispose the relations similar to the ones given by Eqs.
(\ref{EOM-e}) and (\ref{sumrule}) transformed to the mixed
representation (\ref{def3part}). Namely, we have
\begin{eqnarray}
\label{rel-to-e}
e(x) &=& - \frac{1}{2}\frac{d}{dx}
\int_0^1du\, {\scriptstyle\cal Z}(x,u), \\
\label{rel-to-h}
\bar{h}_L(x)&=& - \frac{1}{2}\frac{d}{dx} \int_0^1du\,
(1-2u)\widetilde {\scriptstyle\cal Z}(x,u),
\end{eqnarray}
where we introduce for convenience a new function $\bar{h}_L(x)$,
so that ${h}_L(x)$ reads:
\begin{eqnarray}
\label{sumrule2}
{h}_L(x)=
2 \int_x^1dy\, \frac{x^2}{y^2} h_1(y) -
x\frac{d}{dx} \int_x^1 \frac{dy}{y} \frac{x^2}{y^2} \bar{h}_L(y).
\end{eqnarray}
The last term on the RHS coincides with the twist-3 part
$\widetilde{h}_L$.

Thus we arrive to an explicitly diagonalized form for the structure
functions $f(x)$ (sche\-ma\-ti\-cally):
\begin{equation}
\label{compl-evol-moments}
[f(Q^2)]_n = \int_{0}^{1} du {\cal W}^f(u)
{\scriptstyle\cal Z}^n\left(u,Q^2\right)
= \sum_{l=1}^{n} {\cal J}_n^l \sum_{k=1}^{n} (E^n_{lk})^{-1}
\langle h | {\cal Z}_n^k (Q_0) | h \rangle
\left( \frac{\alpha(Q)}{\alpha(Q_0)} \right)^{-\lambda^l_n/\beta_0}
\end{equation}
where ${\cal W}^f(u)$ is a weight function in Eq. (\ref{rel-to-e})
and (\ref{rel-to-h}) and ${\cal J}_n^l$ is the overlap integral
\begin{equation}
{\cal J}_n^l = \int_{0}^{1}du {\cal W}^f (u) e^n_{l}(u).
\end{equation}

\subsection{Solution of the evolution equations in multicolour QCD.}

The complicated form of the evolution (\ref{compl-evol-moments})
of the twist-3 distributions compels one to look for the simple
approximated solution which could work with reasonable accuracy.
The resolution of this problem is based on the observation that
in the large-$N_c$ limit only the planar diagrams
(Fig. \ref{two-part} (a,d)) survive and the kernel
$P^n_{\scriptscriptstyle\cal Z Z}(u,v)$ has two known dual
eigenfunctions: $1$ and $1 - 2u$ \cite{bbkt96,bel97}, so that
$\int_0^1 du\, e^n_{l}(u) = \delta_{l1} + {\cal O}(1/N_c)$ and
$\int_0^1 du\, (1-2u) e^n_{l}(u) = \delta_{l2} + {\cal O}(1/N_c)$,
where $l=1,2$ correspond to the lowest two eigenvalues of the
spectrum shown in Fig.~\ref{eigenvalues}, thus in the sum in
Eq. (\ref{compl-evol-moments}) only the lowest two terms survive.
Then, a straightforward calculation gives the following DGLAP
evolution kernels:
\begin{equation}
\label{largeNc-ZZ}
\int_0^1 du\,
\!\left\{ \!
\begin{array}{c}
1 \\
\frac{1-2u}{1-2v}
\end{array}
\!\right\}
P_{\scriptscriptstyle\cal Z Z}(x,u,v)
= N_c\,\theta (\bar x)\theta (x)
\!\left\{ \!
\begin{array}{c}
\left[ \frac{x^2}{\bar x} \right]_+ + \frac{1}{2}x^2
- \frac{5}{4} \delta(\bar x) \\
\left[ \frac{x^2}{\bar x} \right]_+ - \frac{3}{2} x^2
- \frac{5}{4} \delta(\bar x)
\end{array}
\!\right\} + {\cal O}\left(\frac{1}{N_c}\right).
\end{equation}
As we have observed previously in section \ref{g_2-evolution}
in the context of the chiral-even distribution $g_2(x)$ \cite{ali91},
similar equations hold true also for the $\frac{1}{N_c}$-suppressed
terms in the $x\to 1$ limit for flavour non-singlet twist-3 evolution
kernels. In the present chiral-odd case, we find
\begin{eqnarray}
\label{largeNc-ZZ-impr}
\int_0^1 du\,
\!\left\{ \!
\begin{array}{c}
1 \\
\frac{1-2u}{1-2v}
\end{array}
\!\right\}
P_{\scriptscriptstyle\cal Z Z}(x,u,v)
=
-\frac{1}{N_c} \theta(\bar x) \theta (x) \!\left\{ \!
\begin{array}{c}
\left[\frac{1}{\bar{x}}\right]_+ + \frac{5}{4}\delta(\bar{x}) +
{\cal O}(\bar{x}^0)
\\
\left[\frac{1}{\bar{x}}\right]_+ + \frac{19}{12}\delta(\bar{x})+
{\cal O}(\bar{x}^0)
\end{array}
\!\right\} + N_c \cdots,
\end{eqnarray}
where the $ N_c \cdots $ symbolize the $x \to 1$ limit of
Eq. (\ref{largeNc-ZZ}).

The eigenfunctions we have obtained coincide precisely with the
coefficients ${\cal W}^f(u)$ that appear in the decomposition
of $e(x,Q^2)$ and $\widetilde{h}_L(x,Q^2)$ in terms of
three-particle correlation functions.

From the observations we have made above, it follows that in the
large-$N_c$ as well as in the large-$x$ limit the twist-3
distributions satisfy the DGLAP (ladder-type) evolution equations
that hold for the twist-2 operators. By combining the large-$N_c$
evolution with the large-$x$ result for the
$\frac{1}{N_c}$-suppressed terms, we can improve the accuracy of
multicolour approximation within a factor 5-10 and reach the
precision of a few per cent as compared with the evolution predicted
using an exact equation (\ref{Mellin-Z}) but supplied with a model
for the light-cone position distribution of gluons between the quark
fields \cite{bel97}. Thus, the functions $e(x,Q^2)$ and
${\bar h}_L(x,Q^2)$ obey the following improved evolution equations:
\begin{eqnarray}
\label{DGLAPforchiodd}
\dot f(x) = {\alpha \over 2\pi} \int_x^1 \frac{dy}{y}
P_f(y)\, f\left( \frac{x}{y} \right),
\end{eqnarray}
with
\begin{eqnarray}
\label{imprKernel}
&&P_{ee}(y) =
2C_F \left[\frac{y}{\bar y}\right]_+ + \frac{C_A}{2}y +
\left(\frac{C_F}{2}-C_A\right)  \delta(\bar y) +
{\cal O}\left({\bar y}^0/N_c\right),\nonumber\\
&&P_{\bar h \bar h}(y)
= 2C_F \left[\frac{y}{\bar y}\right]_+
- \frac{3C_A}{2}y +\left(\frac{7C_F}{6}
-\frac{4C_A}{3}\right)\delta(\bar y)
+{\cal O}\left({\bar y}^0/N_c\right).
\end{eqnarray}
Note that $\widetilde{{h}}_L(x)$ fulfills the same evolution
equation as ${\bar h}_L(x)$. The simplest way to verify this is
to make the Mellin transform of the corresponding evolution
equations.

\subsection{Remarks on the momentum space formalism.}

Let us add a few remarks on the momentum space formulation. As we
have seen above the solution of the evolution equations in the
asymptotic regimes is the most straightforward in the light-cone
position representation. It is by no means trivial to observe
the appearance of the DGLAP equations in momentum fraction
representation. However, we know that the asymptotic solution, in
coordinate space, is given by
the convolution of the three-particle correlation function with
the same weight function that enters in the decomposition of the
two-parton correlators at tree level.
With this in mind, we are able to check that the integrals
\begin{eqnarray}
e(x) &=& \int d\beta' D(x, \beta'),\\
\widetilde h_L (x) &=& x^2 \int_{x}^{1} \frac{d\beta}{\beta^2}
\int \frac{d\beta'}{\beta' - \beta}
\left\{ 2+
(\beta - \beta')
\left[ \frac{\partial}{\partial \beta}
- \frac{\partial}{\partial \beta'} \right]
\right\}
\widetilde D (\beta, \beta'),
\end{eqnarray}
taken from Eqs. (\ref{EOM-e}) and (\ref{sumrule}) neglecting
quark-mass as well as twist-2 effects, satisfy the DGLAP equations,
namely
\begin{eqnarray}
\label{split-N_c-e}
\dot e(x) &=& - \frac{\alpha}{4\pi} N_c \int d \beta e(\beta)
\biggl\{
2
\left[
\frac{ \beta }{( x - \beta )}\Theta_{11}^0 (x, x - \beta)
\right]_+ \nonumber\\
&&+\ \left( 2 - \frac{x}{\beta} \right) \Theta_{11}^0 (x, x-\beta)
- \frac{1}{2} \delta (\beta - x)
\biggr\}, \\
\label{split-N_c-h_L}
\dot{\widetilde h}_L(x) &=& - \frac{\alpha}{4\pi} N_c \int d \beta
\widetilde h_L(\beta)
\biggl\{
2
\left[
\frac{ \beta }{( x - \beta )}\Theta_{11}^0 (x, x - \beta)
\right]_+ \nonumber\\
&&+\ \left( 2 + 3 \frac{x}{\beta} \right) \Theta_{11}^0 (x, x-\beta)
- \frac{1}{2} \delta (\beta - x)
\biggr\}.
\end{eqnarray}
The corresponding anomalous dimensions are
\begin{eqnarray}
\label{e-anomdimesion}
\dot{[e]}_n
&=& \frac{\alpha}{4\pi} N_c
\left\{
-2\psi (n+2) - 2\gamma_E + \frac{1}{2} + \frac{1}{n+2}
\right\}
[e]_n , \\
\dot{[\widetilde h_L]}_n
&=& \frac{\alpha}{4\pi} N_c
\left\{
-2\psi (n+2) - 2\gamma_E + \frac{1}{2} - \frac{3}{n+2}
\right\}
[\widetilde h_L]_n.
\end{eqnarray}
Which are exactly the anomalous dimensions $\gamma^\pm_n$ found in
Ref. \cite{bbkt96} for $e$ and $h_L$, respectively, with the
replacement $n \to j-1$ \footnote{The difference in the anomalous
dimensions is due to an extra power of the momentum fraction $x$
included in the definition of the twist-3 correlation functions.}.

Concluding this section, we have to note that the simple DGLAP
equation (\ref{split-N_c-h_L}) has been used, recently, for
prediction of the size of the structure function $h_L(x)$
\cite{koi97} at high $Q^2$ starting from its value in the low
energy point found in the framework of the MIT bag model. The
main conclusion of this work is that the twist-3 contribution
to $h_L(x)$ is significantly reduced in the course of the evolution
in contrast to the corresponding situation in the case of $g_2(x)$
structure function discussed above. This observation is the 
direct consequence of the larger anomalous dimensions for 
$\widetilde h_L(x)$ at low values of $n$-spins as compared with 
$g_2(x)$. This means that it will be extremely difficult to extract
$\widetilde h_L(x)$ at large momentum transferred. However, if the
latter will be sizable at high $Q^2$ in future experiments it will
indicate that the naive bag model predictions could not be trusted
for the calculations of the quark-gluon correlations presented in
$h_L(x)$ (as well as in the other twist-3 distributions).

\section{Time-like cut vertices.}

\subsection{Time-like processes and the Gribov-Lipatov reciprocity.}

The deep inelastic scattering of lepton beam on the hadron target
has proved to be the most effective experimental tool for studying
the dynamics of hadron reactions on the parton level which has a firm
basis in the quantum field theory provided by the light-cone OPE.
As we have seen in the preceding sections it makes possible the
investigation of the logarithmic violation of the Bjorken scaling as 
well as the power suppressed contributions responsible for polarized
phenomena. However, we mainly use an equivalent approach for the
analysis of the corresponding quantities which is based on the 
factorization theorems and the evolution equations since the
latter can be applicable in the situations when the OPE is no
longer valid. These are the inclusive production of the hadron in
the $e^+e^-$-annihilation, semi-inclusive deep inelastic scattering,
Drell-Yan lepton pair production {\it et ctr}.

There is continuous interest in the inclusive production of
hadrons in hard reactions. These processes involve a quark
fragmentation function to describe the hadron creation from the
underlying hard parton scattering. However, they differ
considerably from the DIS as the short distance expansion could
not be employed although the given processes go near the light cone.
The theoretical basis for strict analyses of the above phenomena
is realized by the generalization of the OPE to the time-like region
in terms of Mueller's $\zeta$-space cut vertices \cite{muel78}. As we
have mentioned in the introduction an essential departure from the
DIS is that the moments of the fragmentation functions are
essentially nonlocal in the coordinate space. However, this
approach has all attractive features of the OPE as it provides
a consistent framework to account for the higher twist effects
\cite{bal91} as well as it allows to sum up the UV logarithms
\cite{cfp80} by using the powerful methods of the renormalization
group.

The semi-inclusive hadron production from a quark fragmentation
is described in QCD by the specific nonperturbative correlation
functions of quark and gluon field operators over the hadron
states which can be identified with $\zeta$-space cut vertices. While
the behaviour of the latters with respect to the fraction of the
parton momentum carried by the hadron is determined by the
nonperturbative strong interaction dynamics, the large $Q^2$-scale
de\-pen\-dence is governed by perturbation theory only. Since the cross
section we are interested in cannot be related to the imaginary part
of some $T$-product of currents, therefore, we must deal with
particular discontinuities of the Feynman diagrams from the very
beginning.

Inasmuch as the twist-3 fragmentation functions enter into several
cross sections on the same footing as the distributions, their scale
dependence is of great interest. Apart from significance for
phenomenology, it is important for theoretical reasons: while it
is know that in the leading order of the coupling constant the
splitting functions for the twist-2 fragmentation functions can be
found from the corresponding space-like quantities via the
Gribov-Lipatov reciprocity relation \cite{grlip72_2}:
\begin{equation}
\label{GL-reciprocity}
P^{\rm SL}\left( x \right) = P^{\rm TL}\left( \frac{1}{x} \right),
\end{equation}
(note that both quantities are defined in the physical regions of
the corresponding channels) no such equality is known for higher
twists.

We begin our discussion with a study of twist-3 nonpolarized
chiral-odd (NCO) fracture functions. This is the simplest case with
respect to the number of correlation functions involved in mixing
under renormalization group evolution. From the phenomenological
point of view they appear, for example, in the cross section for
semi-inclusive hadron ($H$) production in the process of measuring
the nucleon's ($h$) transversity distribution $h_1(x)$ from deep
inelastic scattering \cite{jaf93}:
\begin{eqnarray}
\frac{d^4\Delta\sigma}{dxdyd(1/\zeta )d\phi}
&=&\frac{4\alpha^2_{em}}{Q^2}
\Bigl[
\cos \chi \left( 1-\frac{y}{2} \right)G_1(x,\zeta)\nonumber\\
&+& \cos\phi\sin\chi \sqrt{(\kappa-1)(1-y)}
\left( G_T (x,\zeta) - G_1(x,\zeta)\left( 1-\frac{y}{2} \right) \right)
\Bigr].
\end{eqnarray}
Here $\kappa = 1 + 4 x^2M^2/Q^2$, $y=1-E'/E$ and
the cross section is expressed in a frame where the lepton beam
with energy $E$ defines the $z$-axis and the $x-z$-plane contains
the nucleon polarization vector, which has the polar angle $\chi$
and the scattered electron $E'$ has the polar angles $\theta$,
$\phi$. The functions $G_1$ and $G_T$ are expressed in terms of
the product of the familiar distribution and fragmentation functions
in the following way
\begin{eqnarray}
G_1(x,\zeta)&=&\frac{1}{2}\sum_{i}Q_i^2 g_1^i(x){\cal F}^i(\zeta),\\
G_T(x,\zeta)&=&\frac{1}{2}\sum_{i}Q_i^2
\left[ g_T^i(x){\cal F}^i(\zeta)
+\frac{1}{x}h_1^i(x){\cal I}^i(\zeta) \right].
\end{eqnarray}
All of them have the expressions in QCD in terms of the light-cone
Fourier transformation of correlation functions of fundamental quark
and gluon fields over specific hadron states \cite{jaf91,jaf93}.
Some of the definitions were given already by Eqs. (\ref{g_1&g_2}),
while the others are
\begin{eqnarray}
S^\perp_\mu h_1(x)&=&\frac{1}{2}\int \frac{d\lambda}{2\pi}e^{i\lambda x}
\langle h |\bar\psi (0)i\sigma^\perp_{\mu +}\gamma_5\psi (\lambda n)
|h \rangle, \nonumber\\
{\cal F}(\zeta)&=&\frac{1}{4\zeta}
\int \frac{d\lambda}{2\pi}
e^{i\lambda\zeta}
\langle 0 |
\gamma_+
\psi (\lambda n) |H,X \rangle
\langle H,X|\bar \psi (0) |0 \rangle .
\end{eqnarray}
Note that we have used slightly different definition of the
function $h_1(x)$ which has another dimension in mass units as
compared to previous one. This is done in order to make the product
$h_1(x){\cal I}(\zeta)$ dimensionless. Of course, they can be made
dimensionless separately by introducing certain characteristic
scale $m_{\rm char}^2$ (which can be set equal to the mass $M$ of
the appropriate hadron) into the definition of the corresponding
correlators. ${\cal I}$ is given by equation (\ref{i}).
Note, that the physical regions are different for distribution
and frag\-men\-ta\-tion functions: $0\leq x\leq 1$ and 
$1\leq\zeta<\infty$, respectively.

In the following sections we address ourselves to the problem of
building of the master equation for the function ${\cal I}$.
This problem is of the same complexity as for corresponding
quantities in the space-like domain since we face the mixing with
other cut vertices of the same twist and quantum numbers in the
course of the renormalization group evolution.

\subsection{Recombination functions and their support properties.}

For our purposes it is much more suitable to deal with correlation
functions listed below which are the generalization of the formulae
given in section \ref{e-distribution} to the fragmentation region.

\begin{figure}[htb]
\mbox{
\hspace{0.4cm}
\epsffile{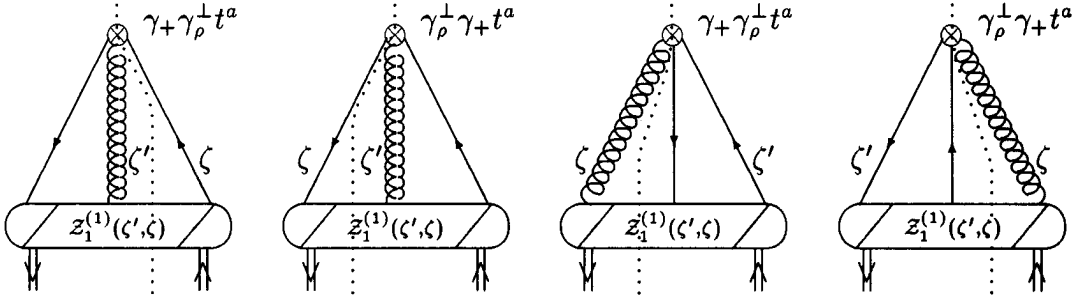}}
\vspace{0.1cm}
{\caption{\label{definition-three-cut}
Graphical representation of the three-parton recombination
functions.
}}
\end{figure}

Diagrammatically, the three-parton correlators are presented in
Fig.~\ref{definition-three-cut}.
\begin{eqnarray}
\label{DefUnpoCF-I}
&&\hspace{-0.7cm}{\cal I}(\zeta)
=\frac{1}{4}\int \frac{d\lambda}{2\pi}
e^{i\lambda\zeta}\label{i}
\langle 0 |
\psi (\lambda n) |H,X \rangle
\langle H,X|\bar \psi (0) |0 \rangle ,\\
&&\hspace{-0.7cm}{\cal M}(\zeta)
=\frac{1}{4\zeta}\int \frac{d\lambda}{2\pi}
e^{i\lambda\zeta}
\langle 0 |
m\gamma_+
\psi (\lambda n) |H,X \rangle
\langle H,X|\bar \psi (0) |0 \rangle ,\\
&&\hspace{-0.7cm}{\cal Z}_1^{(1)}(\zeta',\zeta)
=\frac{1}{4\zeta}\int \frac{d\lambda}{2\pi} \frac{d\mu}{2\pi}
e^{i\lambda\zeta-i\mu\zeta'}
\langle 0 | {\rm g}\gamma_+ \gamma^\perp_\rho
\psi (\lambda n) |H,X \rangle
\langle H,X|\bar \psi (0) B^\perp_\rho (\mu n)|0 \rangle ,\\
&&\hspace{-0.7cm}{\cal Z}_{1}^{(2)}(\zeta,\zeta')
=\frac{1}{4\zeta}\int \frac{d\lambda}{2\pi} \frac{d\mu}{2\pi}
e^{i\mu\zeta'-i\lambda\zeta}
\langle 0 |
{\rm g}\gamma_\rho^\perp\gamma_+ B^\perp_\rho (\mu n)
\psi (0) |H,X \rangle
\langle H,X|\bar \psi (\lambda n) |0 \rangle ,\\
&&\hspace{-0.7cm}{\cal Z}_2^{(1)}(\zeta,\zeta')
=\frac{1}{4\zeta}\int \frac{d\lambda}{2\pi} \frac{d\mu}{2\pi}
e^{i\mu\zeta'-i\lambda\zeta}
\langle 0 |
\bar \psi (0)
{\rm g}\gamma_+ \gamma_\rho^\perp \psi (\mu n) |H,X \rangle
\langle H,X| B^\perp_\rho (\lambda n) |0 \rangle ,\\
&&\hspace{-0.7cm}{\cal Z}_2^{(2)}(\zeta',\zeta)
=\frac{1}{4\zeta}\int \frac{d\lambda}{2\pi} \frac{d\mu}{2\pi}
e^{i\lambda\zeta-i\mu\zeta'}
\langle 0 | B^\perp_\rho (\lambda n) |H,X \rangle
\langle H,X| \bar \psi (\mu n)
{\rm g}\gamma_\rho^\perp \gamma_+ \psi (0) |0 \rangle .\label{last}
\end{eqnarray}
The summation over $X$ is implicit and covers all possible hadronic
final states populated by the quark fragmentation. Again we use the
light-cone gauge $B_+ = 0$, otherwise a link factor should be
inserted in between the quark fields to maintain the gauge
invariance. The quantities determined by these equations form
the closed set under renormalization, however, they are not
independent since there is relation between them due to equation
of motion for the Heisenberg fermion field operator
\begin{equation}
\label{eqmot}
{\cal I}(\zeta)-{\cal M}(\zeta)
-\int d\zeta' {\cal Z}_1(\zeta',\zeta)=0.
\end{equation}
Here and in the following discussion we introduce the convention
\begin{eqnarray}
&&{\cal Z}_j (\zeta',\zeta)
=\frac{1}{2}
\left[ {\cal Z}_j^{(1)}(\zeta',\zeta)+
{\cal Z}_j^{(2)}(\zeta,\zeta')
\right].
\end{eqnarray}
While the former two functions ${\cal I}$ and ${\cal M}$ can be
made explicitly gauge invariant by inserting the $P$-ordered
exponential (which is unity in the gauge we have chosen) between
the quark fields, the latter can be written in the gauge invariant
manner by introducing the following objects:
\begin{eqnarray}
{\cal R}_1(\zeta',\zeta)=\zeta' {\cal Z}_1(\zeta',\zeta),
\hspace{0.8cm}
{\cal R}_2(\zeta,\zeta')=\zeta {\cal Z}_2(\zeta ,\zeta').
\end{eqnarray}
Taking into account Eq. (\ref{B_to_G}) it is easy to verify, in the
same way as before, that they are indeed expressed in terms of
correlators involving the gluon field strength tensor. The
functions ${\cal Z}^{(1)}$ and ${\cal Z}^{(2)}$ are related by
complex conjugation
\begin{eqnarray}
&&\left[{\cal Z}_1^{(1)}(\zeta',\zeta)\right]^*
={\cal Z}_1^{(2)}(\zeta,\zeta'),\hspace{0.5cm}
\left[{\cal Z}_2^{(1)}(\zeta,\zeta')\right]^*
={\cal Z}_2^{(2)}(\zeta',\zeta).
\end{eqnarray}
Their support properties can be found by applying Jaffe's recipe
\cite{jaf83}. It has been shown that the field operators
entering the definition of the correlation functions can be
placed in the arbitrary order on the light cone with appropriate
sign change according to their statistics. Then taking
the particular ordering and saturating the correlation function by
the complete set of the physical states we immediately obtain
(for definiteness, we consider the function ${\cal Z}_1^{(1)}$)
\begin{eqnarray}
{\cal Z}_1^{(1)}(\zeta' ,\zeta)
=\!\!\!\!\!&&\frac{1}{4\zeta}\sum_{X,Y}
\delta (\zeta - 1 -\zeta_X)\delta (\zeta' - \zeta_Y)
\langle 0|\psi |H,X \rangle
\langle H,X | \bar\psi |Y \rangle
\langle Y | B^\perp |0 \rangle\nonumber\\
=&&\!\!\!\!\!\frac{1}{4\zeta}\sum_{X,Y}
\delta (\zeta - 1 -\zeta_X)\delta (\zeta - \zeta' - \zeta_Y)
\langle 0|\psi |H,X \rangle
\langle H,X | B^\perp |Y \rangle \label{pos}
\langle Y | \bar\psi |0 \rangle,
\end{eqnarray}
with $\zeta_X, \zeta_Y \geq 0$ and we omit the unessential Dirac
matrix structure of the vertex. From these equations the
restrictions emerge on the allowed values of the momentum fractions:
$1\leq \zeta <\infty$, $0\leq \zeta'\leq \zeta$. By analogy one can
easily derive similar support properties for other functions.

\subsection{Feynman rules for the discontinuities of the diagrams.
Kel\-dysh diagram technique.}

\begin{figure}[htb]
\mbox{
\hspace{1.2cm}
\epsffile{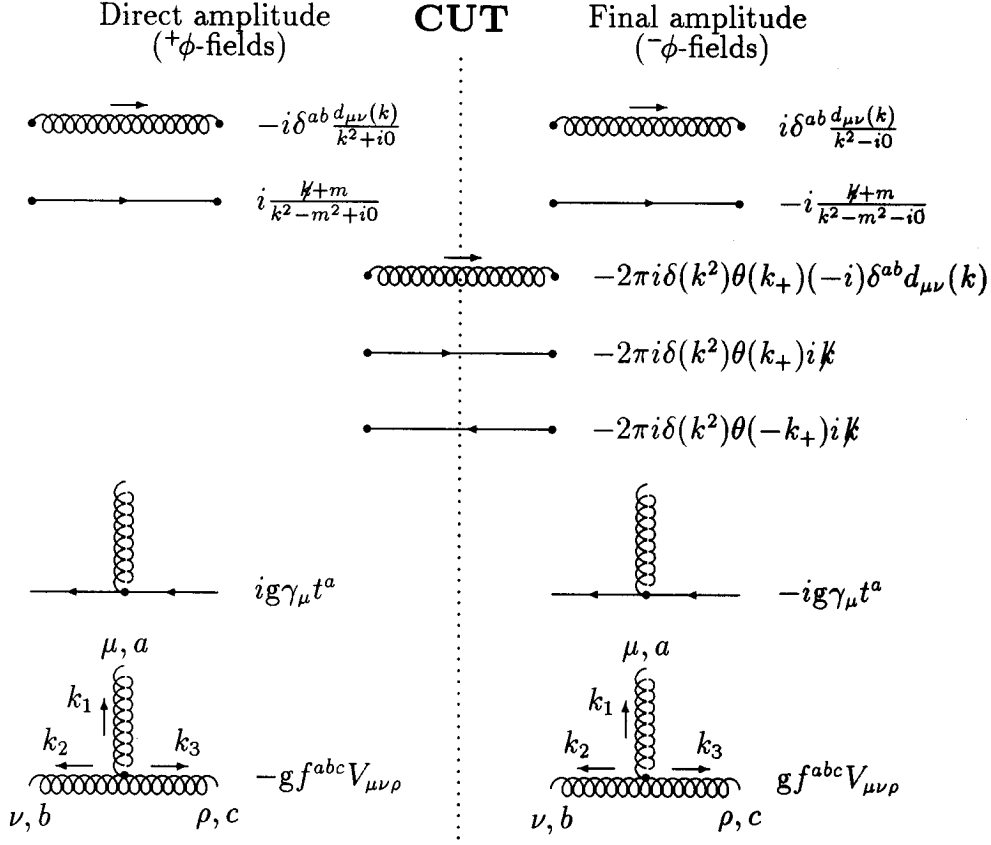}}
\vspace{0.1cm}
{\caption{\label{table}
Modified rules for the discontinuities of the Feynman diagrams.
}}
\end{figure}

It is well known that the time-like cross section could not be
related to the imaginary part of any $T$-product of the currents.
Contrary, it is given by the particular absorptive parts of the
ladder-type diagrams. In the physical region of the annihilation
channel these graphs possess additional discontinuities, which
are not relevant for our purposes, since we are restricted to the
cuts that separate the possible hadrons in the final state. To be
able to extract the imaginary part we are interested in we have
to label in some way the field operators in the amplitudes to
the right and to the left of the cut. This can be suitably done
with the help of the Keldysh diagram technique \cite{kel64}. It
allows to recast the program for the calculation of the particular
discontinuities of the Feynman diagrams to the operator-like
language\footnote{In this discussion we will follow Ref.
\cite{bal91}}. Consider, for instance, certain $S$-matrix
element which is given by the following functional integral
\begin{equation}
M = \int D\phi\ \left( \phi_1 \phi_2 ... \phi_n \right)
\exp\left\{ i \int dz L(\phi) \right\},
\end{equation}
where $\phi = \psi, \bar\psi, B$. The cross section $\sigma =
MM^\dagger $ of the process is
\begin{eqnarray}
MM^\dagger &=& \int D\ {^+\!\phi} D\ {^-\!\phi}\
\left( {^+\!\phi}_1\ {^+\!\phi}_2\ \dots\ {^+\!\phi}_n \right)\
\left( {^-\!\phi}_1\ {^-\!\phi}_2\ \dots\ {^-\!\phi}_n \right)
\nonumber\\
&&\hspace{5cm}\exp\left\{ i \int dz\ {^+\!L}(\phi)
- i \int dz\ {^-\!L}(\phi) \right\}.
\end{eqnarray}
Here the "plus" and "minus" superscripts label the fields from the
direct and conjugated amplitudes. Following the original works
\cite{kel64} one can accept that they are the components of a
unique operator ${\mit\Phi}(z, t)$ composed from the time-
and anti-time-ordered fields, {\it i.e.} $\phi^+$ and $\phi^-$,
respectively. Now, the Green function for the "big" field is a
$2\times 2$-matrix constructed from the usual Feynman propagator,
its conjugated analogue and its discontinuity via the Cutkosky
rules for the lines connecting the direct and final amplitudes.
Thus, the radiative corrections to the bare cut vertex can be
calculated then by using the conventional Feynman rules with
the following modifications:
\begin{itemize}

\item All propagators and vertices on the RHS of the cut are Hermitian
conjugated to that on the LHS.

\item Every time crossing the cut the propagator $1/(k^2-m^2+i0)$ has
to be replaced by $-2\pi i\delta (k^2-m^2)$.

\item For each propagator crossing the cut there is a $\theta$-function
specifying that the energy flow from the LHS to the RHS is positive.
(In the infinite momentum frame this is the plus component of the
four-momentum.)
\end{itemize}
These statements complete the rules to handle the cut vertices. They
can be summarized in the Fig. \ref{table}, where $V_{\mu\nu\rho}
= \left( k_1 - k_2 \right)_\mu g_{\nu\rho}
+\left( k_2 - k_3 \right)_\nu g_{\mu\rho}
+\left( k_3 - k_1 \right)_\rho g_{\mu\nu}$.

\subsection{Abelian evolution.}

In this section accepting the diagram technique derived in the
preceding section we show our machinery on a simple example of
abelian evolution and generalize it afterwards to the Yang-Mills
theory. As in \ref{abel-evol} we start with overcomplete set of
the cut vertices defined by Eqs. (\ref{i})-(\ref{last}) and
disregard for a moment the relation between them. Then
Eq. (\ref{eqmot}) verifies that the evolution equations thus
obtained are indeed correct. We have to note that since the observed
particle $H$ is always in the final state some cuts of Feynman
diagrams are not allowed and, therefore, we could not obtain the
evolution kernel for the cut vertex taking naively the discontinuity
of uncut graph as we are restricted over the limited set of the cuts.

\subsubsection{Sample calculation of the evolution kernels.}

As we have noted previously the UV divergences occur in the
transverse-momentum integrals of partons interacting with a bare
cut vertex. To extract this dependence properly it is sufficient
to separate the perturbative loop from correlation function in
question. To this end, the latter can be represented in the form of
momentum integral in which the integrations over the fractional
energies of the particles attached to the vertex are removed
\begin{equation}
\left(
\begin{array}{c}
{\cal I} (\zeta)\\
{\cal M} (\zeta)
\end{array}
\right)
=
\int \frac{d^4k}{(2\pi)^4}
\delta (\zeta - z)
\left(
\begin{array}{c}
I\\
\frac{1}{\zeta}m\gamma_+
\end{array}
\right)
F(k),
\end{equation}
where
\begin{equation}
F(k)=
\int d^4 xe^{ikx}
\langle 0|
\psi (x)
|H,X\rangle
\langle H,X | \bar\psi (0) |0 \rangle  .
\end{equation}

In the same way we can easily write down the corresponding 
expressions for the three-particle correlation functions.

\begin{figure}[htb]
\mbox{
\hspace{2.4cm}
\epsffile{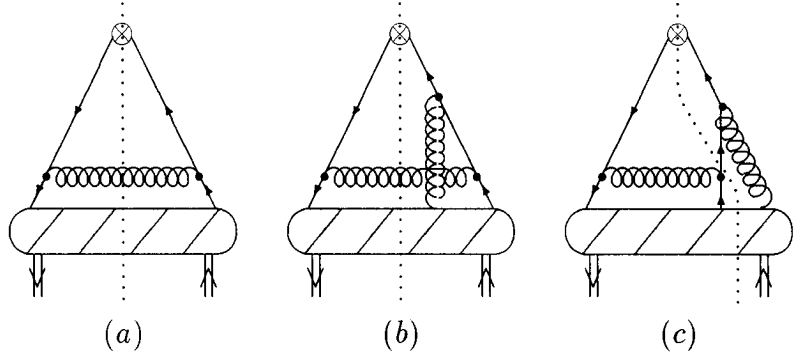}}
\vspace{0.1cm}
{\caption{\label{fig07}
One-loop diagrams contributing to the transition kernels of the
two-particle correlation functions to the two- and three-parton
ones.
}}
\end{figure}

Let us consider, for definiteness, the fracture function ${\cal I}$.
Simple calculation of the one-loop diagram depicted in Fig.
\ref{fig07}~(a) for the $2 \to 2$ transition gives in the LLA
\begin{eqnarray}
{\cal I}(\zeta)_{\Lambda^2}&=&{\rm g}^2
\int\frac{d^4 k}{(2\pi)^4}F(k)
\int \frac{d^4k''}{(2\pi)^3}\delta (\zeta - z'')
\theta (z'' -z)\frac{\delta ((k'' - k)^2)}{k''^4}\nonumber\\
&&\times\{ -d_{\mu \nu } (k'' - k) \gamma_\mu (\not\! k''+m)
I (\not\! k'' +m)\gamma_\nu \}\nonumber\\
&&\nonumber\\
&=& - \frac{\alpha}{2\pi^2}\int \frac{d^4k}{(2\pi)^4}
F(k)\int dz''
\frac{\delta (z'' - \zeta) \theta (z'' -z )}{(z'' - z)}
\int d^2k''_\perp \int d\alpha''
\delta \left(\alpha'' +\frac{k''^2 _\perp}{2(z''-z)}\right)\nonumber\\
&&\times\frac{1}{[2\alpha'' z'' + k''^2_\perp]^2}
\left\{
[2\alpha'' z'' + k''^2_\perp]
- 2m\gamma_+ \left[\alpha''
+\frac{[2\alpha'' z'' + k''^2_\perp]}{(z'' - z)} \right]
\right\}\nonumber\\
&&\nonumber\\
&=&-\frac{\alpha}{2\pi}\ln \Lambda^2
\int \frac{dz}{z}\theta (\zeta - z)
\left[
{\cal I}(z)
-{\cal M}(z)\
\left(
1+2\frac{z}{(\zeta - z)}
\right) \label{two}
\right].
\end{eqnarray}
As long as the logarithmic contribution appears when $|k_\perp|/
|k''_\perp| \ll 1$ and $\alpha / \alpha'' \ll 1$ we expand the
integrand in powers of these ratios keeping the terms that do
produce the logarithmic divergence. Similarly, one can evaluate
the transition amplitudes of ${\cal I}$ to the three-particle
correlation functions ${\cal Z}_j$ (for the diagrammatical
representation, see Figs. \ref{fig07} (b,c)):
\begin{eqnarray}
{\cal I}(\zeta)_{\Lambda^2}&=&
\frac{\alpha}{2\pi}\ln \Lambda^2
\int dz dz' \theta (\zeta - z)
{\cal Z}_1 (z', z)
\left[
\frac{2}{(\zeta - z)}\label{three}
+
\frac{1}{(z-z')},
\right]\\
{\cal I}(\zeta)_{\Lambda^2}&=&
\frac{\alpha}{2\pi}\ln \Lambda^2
\int dz dz' \theta (\zeta - z)
{\cal Z}_2 (z, z')
\frac{(\zeta - z)}{(\zeta -z')(z -z')}.
\end{eqnarray}

\begin{figure}[htb]
\mbox{
\hspace{6.2cm}
\epsffile{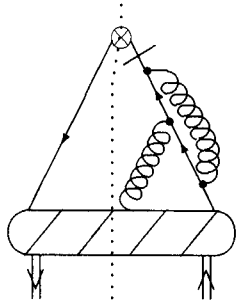}}
\vspace{0.1cm}
{\caption{\label{fig08}
Contact-type contribution to the evolution equation of the
fragmentation function ${\cal I}(z)$.
}}
\end{figure}

Due to the non-quasi-partonic \cite{lip85} form of the vertex $I$
there exists an additional con\-tri\-bu\-tion to the evolution equation
coming from the contact term (Fig. \ref{fig08}) that results from
the cancellation of the propagator adjacent to the quark-gluon and
bare cut vertices. As a consequence the vertex acquires the
three-particle piece
\begin{eqnarray}
{\cal I}(\zeta)_{\Lambda^2}
&=&\int \frac{d^4k}{(2\pi)^4} \frac{d^4k'}{(2\pi)^4}
{\cal Z}_{1\rho}(k',k)
i{\rm g}\Gamma_\rho (k-k', k)i G(k)\delta (\zeta - z) + (c.c.)\\
&=& \frac{\alpha}{2\pi}\ln\Lambda^2
\int dz' {\cal Z}_1(z', \zeta)
\int dz'' \frac{\zeta (\zeta - z'- z'')}{z''}
\Theta^0_{111} (z'' ,z''-\zeta ,z''- \zeta +z').\nonumber
\end{eqnarray}
As can be seen Eqs. (\ref{two}) and (\ref{three}) possess the IR
divergences at $z =\zeta$. They disappear after we account for
the virtual radiative corrections (renormalization of the field
operators) discussed in section \ref{renormalization}. The net
result looks like
\begin{equation}
\Gamma^R = (1 - \Sigma_1)U_1\Gamma U^{-1}_2, \hspace{0.5cm}
\Gamma = \left(I,\hspace{0.2cm} \frac{1}{\zeta}m \gamma_+ ,
\hspace{0.2cm} {\rm g} \gamma_+ \gamma^\perp_\rho  \right)
\label{renorm}
\end{equation}

Assembling all these contributions we come to the evolution equation
for ${\cal I}$ given below by Eq. (\ref{iev}).

\subsubsection{Evolution equations.}

\begin{figure}[htb]
\mbox{
\hspace{0.5cm}
\epsffile{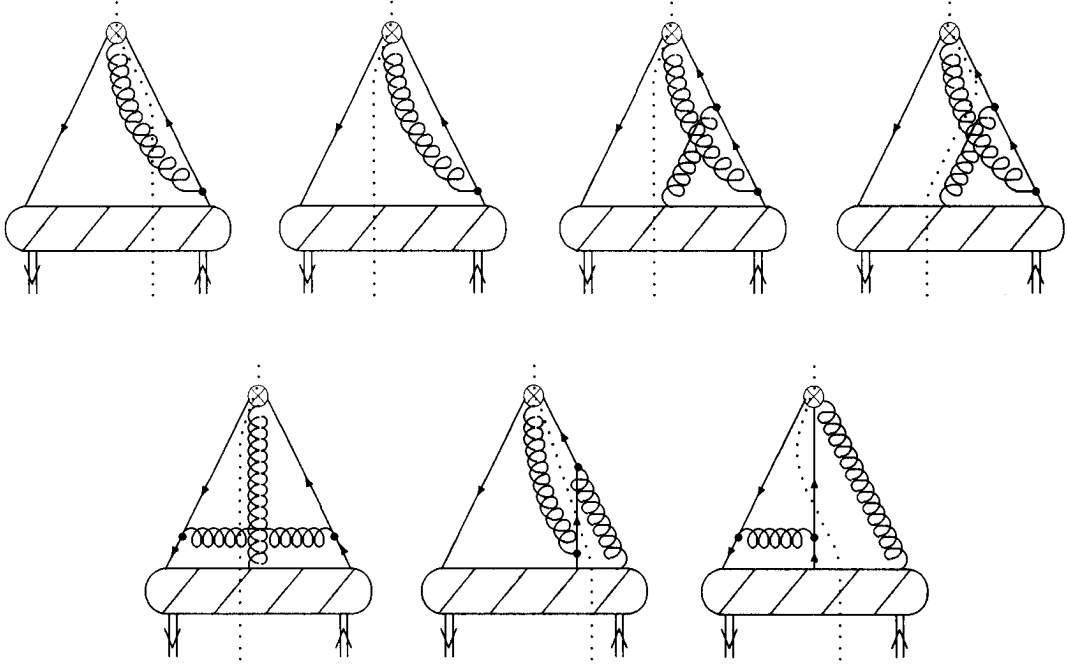}}
\vspace{0.1cm}
{\caption{\label{fig09}
One-loop (abelian) corrections to the three-point correlation
function ${\cal Z}_1$.
}}
\end{figure}

\begin{figure}[htb]
\mbox{
\hspace{0.5cm}
\epsffile{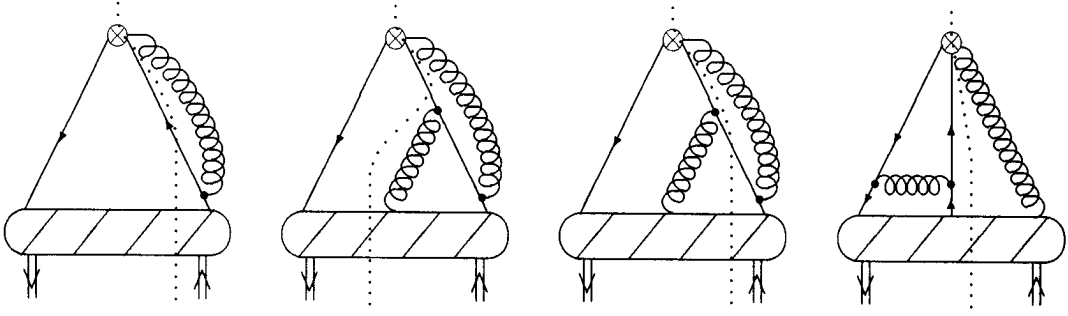}}
\vspace{0.1cm}
{\caption{\label{fig10}
The same as in Fig. \protect\ref{fig09} but for the three-point
correlation function ${\cal Z}_2$.
}}
\end{figure}

Now following the procedure just described it is not difficult to
construct the closed set of the evolution equations by calculating
the one-loop diagrams shown on Figs. \ref{fig09}, \ref{fig10}
\begin{eqnarray}
\label{mass-frag}
\dot{{\cal M}}(\zeta)
&=&\frac{\alpha}{2\pi}\int \frac{dz}{z}\theta (\zeta - z)
P_{\cal MM}\left(\frac{\zeta}{z}\right)
{\cal M}(z),\\
&&\nonumber\\
\label{iev}
\dot{{\cal I}}(\zeta)
&=&\frac{\alpha}{2\pi}\int \frac{dz}{z}\theta (\zeta - z)
\biggl\{
P_{\cal II}\left(\frac{\zeta}{z}\right){\cal I}(z)
+ P_{\cal IM}\left(\frac{\zeta}{z}\right) {\cal M}(z) \\
&+&\int dz'\biggl[
P_{{\cal I} {\cal Z}_1}\left(\frac{\zeta}{z},\frac{z'}{\zeta}\right)
{\cal Z}_1(z',z)
+\frac{z(\zeta - z)}{(\zeta - z')(z - z')}
{\cal Z}_2(z,z')
\biggr]
\biggr\},\nonumber\\
&&\nonumber\\
\label{log}
\dot{{\cal Z}}_1(\zeta',\zeta)&=&
\frac{\alpha}{2\pi}
\biggl\{
\Theta_{11}^0 (\zeta',\zeta'-\zeta)
\left[ {\cal M}(\zeta) -
\frac{(\zeta - \zeta')}{\zeta}{\cal I}(\zeta) \right]\\
&+&
\theta (\zeta')
\left[ \frac{1}{\zeta}{\cal M}(\zeta - \zeta')
- \frac{1}{(\zeta - \zeta')}{\cal I}(\zeta - \zeta')
\right]
\nonumber\\
&+&\int \frac{dz}{z} \theta (\zeta - z)
\biggl[
P_{{\cal Z}_1{\cal Z}_1}
 \left( \frac{\zeta}{z}, \frac{\zeta'}{\zeta} \right)
{\cal Z}_1 (\zeta', z)
+
\frac{z(\zeta - z)^2}{\zeta \zeta' (z - \zeta + \zeta')}
{\cal Z}_2 (z, \zeta - \zeta')
\biggr]\nonumber\\
&+&\int dz' \biggl[
\Theta_{111}^0 (\zeta',\zeta'-\zeta,\zeta'-\zeta+z')
\frac{(\zeta'-\zeta+z')}{\zeta'}
{\cal Z}_1 (z', \zeta)\nonumber\\
&+&
\theta (\zeta')
\frac{(z' - \zeta)}{\zeta (z' - \zeta + \zeta')}
{\cal Z}_1 (z', \zeta - \zeta')
+
\theta (\zeta - \zeta')
\frac{\zeta'(\zeta - \zeta')}{\zeta (\zeta - z')(\zeta' - z')}
{\cal Z}_2 (\zeta',z')
\biggr]
\biggr\},\nonumber\\
&&\nonumber\\
\label{Z^1_2}
\dot{{\cal Z}}_2(\zeta,\zeta')&=&
\frac{\alpha}{2\pi}
\biggl\{
-\theta (\zeta - \zeta')
\biggl[
\frac{1}{\zeta}{\cal M}(\zeta')
-\frac{(\zeta' - \zeta)}{\zeta \zeta'}{\cal I}(\zeta ')
\biggr]\\
&+&\int \frac{dz}{z} \theta (\zeta -z)\frac{z(z-\zeta)}{\zeta^2}
{\cal Z}_1(z-\zeta,z)\nonumber\\
&-&\int dz'
\biggl[
\theta (\zeta -\zeta')
\frac{(\zeta-\zeta')(\zeta-\zeta'+z')}{\zeta^2 z'}
{\cal Z}_1(z',\zeta')\nonumber\\
&+&\Theta^0_{11}(\zeta',\zeta'-z')
\frac{\zeta'}{(\zeta' - z')}
\left[{\cal Z}_2(\zeta,z')
-{\cal Z}_2(\zeta,\zeta')\right]\nonumber\\
&+&\Theta^0_{11} (\zeta' - \zeta, \zeta' - z')
\frac{(\zeta' - \zeta)}{(\zeta' - z')}
\left[
{\cal Z}_2(\zeta,z')
-{\cal Z}_2(\zeta, \zeta')
\right]
\biggr]
+\frac{3}{2}{\cal Z}_2(\zeta, \zeta')
\biggr\}.\nonumber
\end{eqnarray}
where the dot denotes the derivative with respect to the UV cutoff
$\hspace{0.1cm}\dot{}=\Lambda^2{\partial}/{\partial\Lambda^2}$ and
splitting functions are given by the following equations
\begin{eqnarray}
P_{\cal MM}(z)
&=&-\left[\frac{2}{z(1-z)}\right]_+ + \frac{1}{z} + 1,\\
P_{\cal II}(z)
&=& -1 + \frac{1}{2}\delta (z-1),\\
P_{\cal IM }(z)
&=&-\left[\frac{2}{z(1-z)}\right]_+ + \frac{2}{z} + 1, \\
P_{{\cal I} {\cal Z}_1}(z,y)
&=&-\left[\frac{2}{z(1-z)}\right]_+ + \frac{2}{z}
+ \frac{1}{1-yz} - \delta (z-1) \frac{1}{y}\ln (1-y),\\
P_{{\cal Z}_1 {\cal Z}_1}(z,y)
&=&-\left[\frac{2}{z(1-z)}\right]_+ + \frac{2}{z}
+\frac{y}{1-yz}
+ \delta (z-1) \left[ \frac{3}{2} - \ln (1-y) \right].
\end{eqnarray}

Now it is an easy task to verify the fulfillment of the equation of
motion (\ref{eqmot}) for the correlation functions as a consistency
check of our calculations. By exploiting this relation we exclude
${\cal I}$ from the above set of functions and reduce the system to
the basis of independent gauge invariant quantities $\{{\cal M}$ and
${\cal R}_j\}$.

An important note is in order now. As distinguished from the
evolution of the structure functions, the above Eq. (\ref{log}) has
the logarithmic dependence on the ratio of the parton momentum
fractions\footnote{This is consequence of the fact that while the
support properties of the multi-parton recombination functions are
different from the distributions, {\it i.e.} the region of attained
momentum fractions is nonsymmetric in the former case in contrast
to the latter, the perturbative one-loop renormalization of the field
operators is not sensitive to this. Actually, it is given by the same
equations as in the previous discussion of the deep inelastic
scattering}. The consequence of its presence is obvious. Taking into
account the restrictions imposed by Eq. (\ref{pos}) we can define the
moments of the correlation functions in the following way
\begin{eqnarray}
&&{\cal M}_n =\int_{1}^{\infty}
\frac{d\zeta}{\zeta^n}{\cal M}(\zeta),\\
&&{\cal R}^m_n =\int_{1}^{\infty}\frac{d\zeta}{\zeta^n}
\int_{0}^{\zeta} d \zeta' \zeta'^m {\cal R}(\zeta',\zeta).
\end{eqnarray}
We find for two-particle cut vertex
\begin{eqnarray}
&&\dot{\cal M}_n
=- \frac{\alpha}{2\pi} \left( S_n + S_{n-2} \right) {\cal M}_n.
\end{eqnarray}
Comparing it with Eqs. (\ref{comparison-1}) and (\ref{comparison-2})
we notice the universality of the evolution kernels for the space-
and time-like two-particles quasi-partonic cut vertices, {\it i.e.}
the Gribov-Lipatov reciprocity (\ref{GL-reciprocity}) relation, which
looks like
\begin{equation}
\label{GL-anomdim}
\gamma_{n+2}^{\rm TL} = \gamma_n^{\rm SL}
\end{equation}
in terms of the corresponding anomalous dimensions, is fulfilled.

On the other hand, it is impossible to write down the finite system
of equations for any given physical moment of the three-parton 
correlation functions as the logarithm of the ratio of the parton 
momentum fractions in the evolution kernels leads to the infinite 
series of the moments as distinguished from the deep inelastic 
scattering where the rank of the anomalous dimension matrix was 
finite and increases with the number of the moment. This is 
consequence of essential nonlocality of the cut vertices in the 
coordinate space since even if we start from the local cut vertex 
it will smeared along the light cone upon the renormalization. 
Therefore, it is not possible to solve the system of equations 
successively in terms of moments as well as we do not succeed in 
solving it analytically in a general form. However, in the next 
section when dealing with the QCD evolution we will find that the 
system of coupled equations (\ref{log}) and (\ref{Z^1_2}) can be 
reduced to the single equation in the multicolour limit and its 
solution can be found analytically. 

\subsection{Non-abelian evolution.}

\begin{figure}[htb]
\mbox{
\hspace{2.4cm}
\epsffile{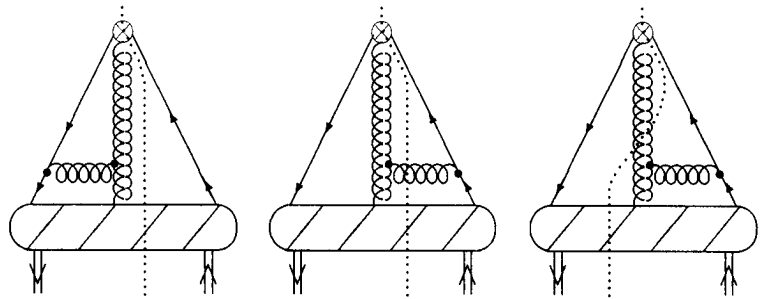}}
\vspace{0.1cm}
{\caption{\label{fig11}
Non-abelian radiative corrections to the evolution kernels of the
fragmentation function ${\cal Z}_1$.
}}
\end{figure}

\begin{figure}[htb]
\mbox{
\hspace{2.4cm}
\epsffile{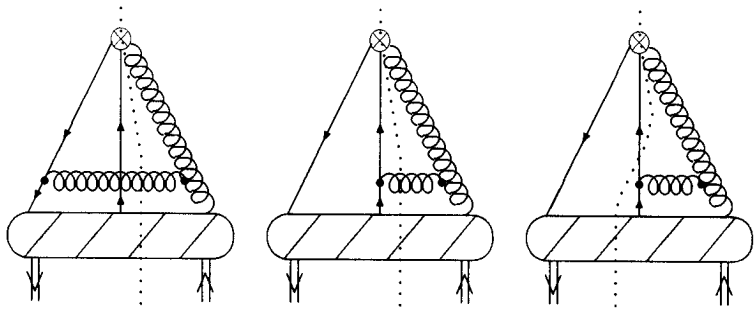}}
\vspace{0.1cm}
{\caption{\label{fig12}
The same as in Fig. \protect\ref{fig11} but for the fragmentation
function ${\cal Z}_2$.
}}
\end{figure}

In the QCD case we should add the diagrams with triple-boson
interaction vertex (see Figs. \ref{fig11}, \ref{fig12}) and gluon
self-energy insertions (not shown).

Gathering these contributions together with equations obtained in the
previous section (with colour group factors accounted for properly),
we come to the final result
\begin{eqnarray}
&&\hspace{-0.7cm}\dot{\cal Z}_1(\zeta',\zeta)=
\frac{\alpha}{2\pi}
\Biggl\{
C_F
\left[
\Theta^0_{11} (\zeta', \zeta' - \zeta)\frac{\zeta'}{\zeta}
{\cal M}(\zeta)
-
\theta (\zeta') \frac{\zeta'}{\zeta (\zeta - \zeta')}
{\cal M}(\zeta -\zeta')
\right]\nonumber\\
&&\hspace{-0.7cm}+\int \frac{dz}{z}\theta (\zeta - z)
\Biggl[
P_{{\cal Z}_1{\cal Z}_1}\left( \frac{\zeta}{z},
\frac{\zeta'}{\zeta} \right)
{\cal Z}_1 (\zeta' ,z)
+ \left( C_F -\frac{C_A}{2} \right)
\frac{z(\zeta - z)^2}{\zeta \zeta' (z - \zeta' + \zeta)}
{\cal Z}_2(z,\zeta - \zeta')\nonumber\\
&&\hspace{-0.7cm}+\ \frac{C_A}{2}
\Biggl(
-2z \frac{\partial}{\partial\zeta'}\int_{0}^{1}dv
{\cal Z}_1(\zeta' - v(\zeta - z), z)
+\left(\frac{z}{(z-\zeta+\zeta')}
-\frac{z(z+\zeta')}{\zeta \zeta'} \right)
{\cal Z}_1(z- \zeta+\zeta',z)
\Biggr)
\Biggr]\nonumber\\
&&\hspace{-0.7cm}+\int dz'
\Biggl[
-C_F
\left(
\Theta^0_{11} (\zeta',\zeta'-\zeta)\frac{(\zeta - \zeta')}{\zeta}
{\cal Z}_1(z',\zeta)
+\theta (\zeta')\frac{1}{(\zeta- \zeta')}
{\cal Z}_1(z',\zeta-\zeta')
\right)\nonumber\\
&&\hspace{-0.7cm}+\left(C_F - \frac{C_A}{2} \right)
\Biggl(
\Theta^0_{111}(\zeta', \zeta' - \zeta, \zeta' - \zeta +z')
\frac{(\zeta' - \zeta +z')}{\zeta'}
{\cal Z}_1(z',\zeta)\nonumber\\
&&\hspace{-0.7cm}+\ \theta (\zeta')
\frac{(z' -\zeta)}{\zeta (z' - \zeta + \zeta')}
{\cal Z}_1(z', \zeta- \zeta')
+ \theta (\zeta - \zeta')
\frac{\zeta' (\zeta - \zeta')}{\zeta (\zeta -z')(\zeta' -z')}
{\cal Z}_2(\zeta', z')
\Biggr)\nonumber\\
&&\hspace{-0.7cm}+\ \frac{C_A}{2}
\Biggl(
\Theta^0_{111}(\zeta', \zeta' -\zeta, \zeta' -z')
\frac{z'(z' +\zeta' - \zeta)}{\zeta' (\zeta' - z')}
{\cal Z}_1(z',\zeta)\nonumber\\
&&\hspace{-0.7cm}+\ \Theta^0_{11}(\zeta', \zeta' - \zeta)
\frac{(\zeta - \zeta')(\zeta' + z')}{(\zeta' - z')(\zeta - z')}
{\cal Z}_1(z',\zeta)
+\Theta^0_{11}(\zeta',\zeta' - z')
\frac{(\zeta'+\zeta)}{(\zeta -z')}{\cal Z}_1(z',\zeta)\nonumber\\
&&\hspace{-0.7cm}-\ \theta (\zeta')
\left( \frac{1}{(z'-\zeta')} - \frac{(\zeta - \zeta' +z')}{\zeta z'}
\right)
{\cal Z}_1(z' -\zeta' , \zeta - \zeta')\nonumber\\
&&\hspace{-0.7cm}-\ 2\Theta^0_{11}(\zeta', \zeta' - z')
\frac{\zeta'}{(\zeta' - z')}
[{\cal Z}_1(z',\zeta)-{\cal Z}_1(\zeta' , \zeta)]
\Biggr)
\Biggr]
\Biggr\},\\
&&\nonumber\\
&&\hspace{-0.7cm}\dot{\cal Z}_2(\zeta, \zeta')=
\frac{\alpha}{2\pi}
\Biggl\{
-C_F \theta(\zeta - \zeta')\frac{1}{\zeta'}{\cal M}(\zeta')\nonumber\\
&&\hspace{-0.7cm}+ \int \frac{dz}{z}\theta (\zeta - z)
\Biggl[
P_{{\cal Z}_2{\cal Z}_2}
\left( \frac{\zeta}{z},\frac{\zeta'}{\zeta} \right)
{\cal Z}_2(\zeta ,z)
+ \left( C_F -\frac{C_A}{2} \right)
\frac{z(z-\zeta)}{\zeta^2}{\cal Z}_1(z -\zeta,z)\nonumber\\
&&\hspace{-0.7cm}+\ \frac{C_A}{2}
\Biggl(
-2z\frac{\partial}{\partial\zeta'}\int_0^1 dv
{\cal Z}_2 (z,\zeta' -v(\zeta -z))
+\left(
\frac{z}{(z-\zeta +\zeta')}-\frac{z(z+\zeta)}{\zeta^2}
\right)
{\cal Z}_2(z, z-\zeta+\zeta')
\Biggr)
\Biggr]\nonumber\\
&&\hspace{-0.7cm}- \int dz'
\Biggl[
C_F\theta (\zeta - \zeta')\frac{\zeta - \zeta'}{\zeta \zeta'}
{\cal Z}_1(z', \zeta')
\nonumber\\
&&\hspace{-0.7cm}+ \left(C_F - \frac{C_A}{2} \right)
\Biggl(
\theta (\zeta -\zeta')
\frac{(\zeta - \zeta')(\zeta - \zeta'+z')}{\zeta^2 z'}
{\cal Z}_1(z', \zeta')\nonumber\\
&&\hspace{-0.7cm}+\ \Theta^0_{11} (\zeta' , \zeta'-z')
\frac{\zeta'}{(\zeta' - z')}
[{\cal Z}_2(\zeta ,z')-{\cal Z}_2(\zeta , \zeta')]
\nonumber\\
&&\hspace{-0.7cm}+\ \Theta^0_{11}(\zeta' - \zeta, \zeta' - z')
\frac{(\zeta' - \zeta)}{(\zeta' -z')}
[{\cal Z}_2(\zeta, z')-{\cal Z}_2(\zeta ,\zeta')]
\Biggr)\nonumber\\
&&\hspace{-0.7cm}-\ \frac{C_A}{2}
\theta (\zeta -\zeta')
\frac{(\zeta - \zeta')}{(\zeta - z')}
\left(
\frac{1}{(\zeta' - z')}
+\frac{z'}{\zeta^2}
\right)
{\cal Z}_1(z',\zeta')
\Biggr]
\Biggr\},
\end{eqnarray}
where
\begin{eqnarray}
P_{{\cal Z}_1{\cal Z}_1}(z,y)
&=&C_F \left\{
-\left[\frac{2}{z(1-z)} \right]_+ +\frac{2}{z}
+\delta(z-1)\left[ \frac{3}{2} -\ln (1-y)\right]
\right\}\nonumber\\
&+&\left( C_F -\frac{C_A}{2}\right)
\frac{y}{1-yz},\nonumber\\
P_{{\cal Z}_2{\cal Z}_2}(z,y)
&=&\frac{C_A}{2}
\left\{
-\left[
\frac{4}{z(1-z)}
\right]_+
+\frac{4}{z}
+\frac{1}{1-yz}
-\frac{1+z}{z^2}
\right\}\nonumber\\
&+&\delta (z-1)
\left[
\frac{3}{2}C_F
-\frac{C_A}{2}
(\ln y + \ln (1-y ))
\right].
\end {eqnarray}
These equations should be supplemented by the equation for the mass
dependent correlator ${\cal M}$ which differs from its abelian
analogue (\ref{mass-frag}) only by the Casimir operator $C_F$.

\subsection{Asymptotic solution of the evolution equations.}

One can easily observe the significant reduction of the above
evolution equations if we neglect the terms in the kernel of the
order of magnitude ${\cal O}(1/N_c^2)$. In this case an additional
three-parton correlator
${\cal Z}_2 \sim \langle 0 | \bar\psi \psi |H,X \rangle
\langle H,X| B^\perp |0 \rangle $, which appears only through the
radiative corrections, decouples from the evolution equation for
${\cal Z}_1$. Therefore, discarding the quark mass cut vertex we
obtain homogeneous equation which governs the $Q^2$-dependence of
the three-parton correlation function ${\cal Z}_1$.

The situation has the closer similarity with phenomenon found in the
evolution equations for chiral-even and -odd distribution functions
discussed in the first part of this paper \cite{ali91,bbkt96,bel97}
where in the multicolour limit ($N_c \to \infty$) there was a very
important simplification as the evolution kernels have been vanishing
for contributions with interchanged order of partons on the light
cone, {\it i.e.} the momentum fraction carried by gluon in the matrix
element of quark-gluon correlator varies only among the quark ones
and does not exceed the latters. This property allowed to find
the solution of simplified equations exactly in the nonlocal form.
In the present case the decoupling of ${\cal Z}_1$ has the same
consequences.

In the large-$N_c$ limit the RG equation takes the form
\begin{equation}
\dot{\cal Z}_1(\zeta', \zeta)
= \frac{\alpha}{4\pi}
\int dz' \frac{dz}{z} \theta (\zeta - z)
{\cal K} (z, z', \zeta, \zeta') {\cal Z}_1(z', z)
\end{equation}
and the evolution kernel is given by the following expression:
\begin{eqnarray}
\frac{1}{N_c}{\cal K} (z, z', \zeta, \zeta')
&=& 2 \frac{z}{\zeta} \delta (\zeta' - \zeta + z - z')
- \delta (\zeta' - \zeta + z) \\
&&\hspace{-3cm} - \,\frac{2}{\frac{\zeta}{z}
\left( 1 - \frac{\zeta}{z}\right)}
\delta (\zeta' - \zeta + z - z')
+ 2 \int_{1}^{\infty} \frac{dz''}{z''(1-z'')}
\delta \left( 1 - \frac{\zeta}{z} \right)
\delta (\zeta' - z') \nonumber\\
&&\hspace{-3cm} + \,\left[
\delta (\zeta' - \zeta + z - z')
- \delta (\zeta' - \zeta + z)
\right]
\left[
\frac{z}{z'} - \frac{z(z' + \zeta)}{\zeta (z' - z + \zeta)}
\right] \nonumber\\
&&\hspace{-3cm} + \,\delta \left( 1 - \frac{\zeta}{z} \right)
\biggl\{
\frac{3}{2} \delta (\zeta' - z')
- \ln \left( 1 - \frac{z'}{z} \right)\delta (\zeta' - z')
- 2 \left[ \frac{z'}{\zeta' - z'}
\Theta^0_{11} (\zeta', \zeta' - z')\right]_+ \nonumber\\
&&\hspace{-3cm} + \,\frac{z - \zeta'}{\zeta' - z'}
\left[
\frac{\zeta'}{z - z'} + \frac{z'}{\zeta'}
-\frac{\zeta' - z'}{z}
\right]
\Theta^0_{11} (\zeta', \zeta' - z)
+ \left[
\frac{\zeta'}{z - z'} + \frac{z'}{\zeta'} - 1
\right]
\Theta^0_{11} (\zeta', \zeta' - z')
\biggr\}.\nonumber
\end{eqnarray}

Inspired by our knowledge acquired from the previous study of the
twist-3 structure functions we are able to check that Eq. (\ref{eqmot})
(with ${\cal M}=0$) satisfies the ladder-type evolution equation with
the following splitting function:
\begin{equation}
\label{ker}
\frac{1}{N_c}\int d\zeta' {\cal K} (z, z', \zeta, \zeta')
= - \frac{2}{\left[ \frac{\zeta}{z}
\left( 1 - \frac{\zeta}{z}\right) \right]_+}
+ 2 \frac{z}{\zeta} - 1 + \frac{1}{2}
\delta \left( 1 - \frac{\zeta}{z} \right).
\end{equation}
Thus, for the moments we obtain the following solution of the RG
equation ($Q>Q_0$):
\begin{equation}
\label{moments}
\int_{1}^{\infty} \frac{dz}{z^n} {\cal I} (z, Q)
= \left( \frac{\alpha(Q)}{\alpha(Q_0)} \right)
^{^{\rm NCO}\gamma_n/\beta_0}
\int_{1}^{\infty} \frac{dz}{z^n} {\cal I} (z, Q_0),
\end{equation}
and the corresponding anomalous dimensions equal
\begin{equation}
\label{NCO}
{^{\rm NCO}\gamma}_n = N_c
\left\{
- 2 \psi (n-1) - 2 \gamma_E - \frac{3}{n-1} + \frac{1}{2}
\right\},
\end{equation}
as usual $\beta_0 = \frac{2}{3}N_f - \frac{11}{3}C_A$.

As we have previously mentioned, there exists an equation which
states that in the leading log approximation the time-like (TL)
and space-like (SL) kernels corresponding to the twist-2
parton densities are directly related by the Gribov-Lipatov
equations (\ref{GL-reciprocity}) or (\ref{GL-anomdim}). Comparing
the result given by Eq. (\ref{NCO}) with the large-$N_c$ anomalous
dimensions (\ref{e-anomdimesion}) we see the absence of the
universality of the corresponding twist-3 evolution kernels,
{\it i.e.} the Gribov-Lipatov reciprocity is violated.

\subsection{Generalization to other fragmentation functions.}

\begin{figure}[htb]
\mbox{
\hspace{6.2cm}
\epsffile{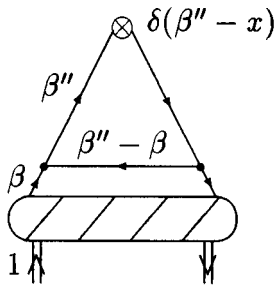}}
\vspace{0.1cm}
{\caption{\label{fig13}
One-loop ladder-type diagram for the two-particle evolution kernel.
}}
\end{figure}

Now we can proceed further and demonstrate that the evolution kernels
for the time-like two-quark densities can directly be found from their
space-like analogues by exploiting the particular form of the evolution
equations given by Eqs. (\ref{split-N_c-e}) and (\ref{split-N_c-h_L}).
Since the analytic structure of the uncut diagram (see Fig. \ref{fig13})
is completely characterized by the integral representation of the
$\Theta$-function given by Eq. (\ref{Theta}), we can just take its
particular discontinuities, using the usual Cutkosky rules supplied
with appropriate theta-function specifying the positivity of the
energy flow from the right- to the left-hand side of the cut, in order
to obtain the corresponding time-like kernel. Since the observed
particle is always in the final state for the fragmentation process,
we are restricted to the single cut\footnote{For a given graph the
possible cuts correspond to the possible final states.} across the
horizontal rank of the ladder diagram in Fig. \ref{fig13}. Namely,
using the integral representation of the corresponding step function
$\Theta^0_{11}$, we have
\begin{equation}
\Theta^0_{11}
(x,x - \beta)=\int_{-\infty}^{\infty}\frac{d\alpha}{2\pi i}
\frac{1}{[\alpha x - 1 +i0][\alpha (x - \beta) - 1 +i0]}
\stackrel{\rm disc}{\longrightarrow}
-\frac{\theta (x - \beta)}{\beta}.
\end{equation}
The self-energy insertions are not affected by the cut since it
does not cross the corresponding lines. Taking into account
different kinematic definitions of the correlation functions
in the space- and time-like regions\footnote{See, for instance
Eqs. (\protect\ref{DefUnpoCF-e}) and (\protect\ref{DefUnpoCF-I}).}
\cite{col82}, we are able to find the kernels. It is easy to
verify that the evolution kernels constructed for the time-like
twist-2 cut vertices using this recipe coincide with the known
results. In the same way, we may obtain the above equation
(\ref{ker}) from Eq. (\ref{split-N_c-e}).

Since there exist fragmentation functions corresponding to each
distribution, apart from the specific ones appearing from the
final state interaction, we are in a position to find large-$N_c$
anomalous dimensions, which govern their $Q^2$-dependence, from
the results (\ref{split-N_c-h_L}) and (\ref{split-N_c-g_2}). Namely,
the genuine twist-3 contributions ${\cal H}^{\rm tw-3}_L$ (PCO)
and ${\cal G}^{\rm tw-3}_T$ (PCE) to the corresponding fragmentation
functions
\begin{eqnarray}
&&\hspace{-0.7cm}{\cal H}_L(\zeta)
=\frac{1}{4}\int \frac{d\lambda}{2\pi}
e^{i\lambda\zeta}
\langle 0 |
i \sigma _{+-} \gamma_5 \psi (\lambda n) |H,X \rangle
\langle H,X|\bar \psi (0) |0 \rangle ,\\
&&\hspace{-0.7cm}{\cal G}_T(\zeta)
=\frac{1}{4}\int \frac{d\lambda}{2\pi}
e^{i\lambda\zeta}
\langle 0 |
\gamma_\perp \gamma_5 \psi (\lambda n) |H,X \rangle
\langle H,X|\bar \psi (0) |0 \rangle .
\end{eqnarray}
after subtracting out the twist-2 piece \cite{bal91} obey the
evolution equation (\ref{moments}) with the following anomalous
dimensions:
\begin{eqnarray}
{^{\rm PCO}\gamma}_n &=& N_c
\left\{
- 2 \psi (n-1) - 2 \gamma_E + \frac{1}{n-1} + \frac{1}{2}
\right\},\\
{^{\rm PCE}\gamma}_n &=& N_c
\left\{
- 2 \psi (n-1) - 2 \gamma_E - \frac{1}{n-1} + \frac{1}{2}
\right\}.
\end{eqnarray}
Of course, it is a trivial task to invert the moments and to find
the DGLAP kernels themselves.

To summarize this section, we have found that in the multicolour
limit of QCD the twist-3 fragmentation functions obey the
ladder-type evolution equations and the corresponding anomalous
dimensions are known analytically. The Gribov-Lipatov reciprocity
is not the property of the twist-3 distributions but it is strongly
violated already in the LLA of the perturbation theory.

\section{Discussion and conclusion.}

We review above the approach to an analysis of the logarithmic
violation of the Bjorken scaling in the twist-3 distribution and
fragmentation functions of the nucleon. It consists in the studying
of the one-loop renormalization of the multi-parton correlators
which explicitly involve the gluon degrees of freedom and further
reconstruction of the evolution equation for them in the LLA using
the renormalization group invariance. For these purposes, we have
used the techniques, which employ the light-like gauge for the gluon
field. The physically transparent picture which appears in this gauge
(which is an essential ingredient of our method) makes the calculations
simple. Accepting different prescriptions on the spurious IR pole in
the gluon propagator, we were able to verify that they do lead to the
same results. We present an exact leading-order evolution for the
correlators in the light-cone fraction as well as in the light-cone
position representations, which display the complementary aspects of
the factorization, and establish the bridge between different
formulations of the QCD evolution.

From the calculational point of view the momentum space technique
is much easier to treat. However, the coordinate space makes the
involved symmetries apparent and, as a by-product, diagonalization
of evolution kernels is easy to handle. The complicated form of exact
master equations for the twist-3 functions compels one to look for
the approximation which could work with reasonable accuracy. The
solution of this problem has been found in the fact that in the
multicolour limit of QCD they are reduced to the ladder-type evolution
equations which generally have very good precision, at the level of
few per cent.

Comparing the analytical expressions for the anomalous dimensions of
the twist-3 struc\-tu\-re and fragmentation functions we observe that the
Gribov-Lipatov reciprocity relation, valid for low-twist parton
densities in the LLA, is broken already in the leading-order of
per\-tur\-ba\-tion theory.

The result we have discussed here are important from the theoretical
point of view since they enrich the theory of the higher-twist
effects in the hadron reactions as well as for phenomenology and
could be used to analyze the experimental data when these become
available.

\bigskip

{\bf Acknowledgments.} We would like to thank the organizers of the
XXXI PNPI Winter School for their hospitality and to D.I.~Diakonov
for the opportunity to deliver this lecture. We are grateful to
D.~M\"uller for fruitful collaboration on the twist-3 evolution.
This work was supported by Russian Foundation for Fundamental
Research, grant N 96-02-17631.

\end{document}